\documentclass[runningheads,envcountsame,envcountsect]{llncs}

\newif\iffull
\fulltrue
\newcommand{\full}[2]{\iffull#1\else#2\fi}
\newif\ifanon
\iffull\anonfalse\else\anontrue\fi

\usepackage{amsmath}
\usepackage{amssymb}
\usepackage{graphicx}
\usepackage{textcomp}
\usepackage{xcolor}
\usepackage{array}
\usepackage[section]{placeins}
\usepackage{tikz}
\usetikzlibrary{arrows.meta,positioning}
\usepackage{tcolorbox}
\usepackage[nospace]{cite}
\usepackage[hidelinks]{hyperref}
\usepackage[expansion=false]{microtype}
\AtBeginDocument{\def\doi#1{\href{https://doi.org/#1}{\nolinkurl{https://doi.org/#1}}}}

\newcommand{\mypara}[1]{\smallskip\noindent\textbf{#1.}}
\newcommand{\temph}[1]{\textbf{#1}}
\newcommand{\App}[1]{\full{Section}{Appendix}~\ref{#1}}

\ifanon
\newcommand{\repo}{\url{https://anonymous.4open.science/r/GLP-02A3}}
\else
\newcommand{\repo}{\url{https://github.com/EShapiro2/GLP}}
\fi

\newcommand{\ia}{\textit{i}}
\newcommand{\ib}{\textit{ii}}
\newcommand{\ic}{\textit{iii}}
\newcommand{\id}{\textit{iv}}
\newcommand{\iie}{\textit{v}}

\newtcolorbox{eg}{colback=gray!5!white,colframe=black!75!black,top=2pt,bottom=2pt}
\usepackage{enumitem}
\setlist{nosep, leftmargin=*}

\title{Grassroots Currencies:\\ Turning Mutual Trust into Liquidity\full{\\ (Full Version)}{}}
\titlerunning{Grassroots Currencies}
\ifanon
\author{Anonymous Submission}
\authorrunning{Anonymous Submission}
\institute{}
\else
\author{Ehud Shapiro}
\authorrunning{E. Shapiro}
\institute{London School of Economics, UK, and Weizmann Institute of Science, Israel}
\fi

\makeatletter
\g@addto@macro\normalsize{%
  \abovedisplayskip 1\p@ \@plus 1\p@ \@minus 1\p@
  \belowdisplayskip \abovedisplayskip
  \abovedisplayshortskip 1\p@ \@plus 1\p@
  \belowdisplayshortskip 3\p@ \@plus 1\p@ \@minus 1\p@}
\def\@spthm#1#2#3#4{\topsep 1\p@ \@plus 1\p@ \@minus 1\p@
\refstepcounter{#1}%
\@ifnextchar[{\@spythm{#1}{#2}{#3}{#4}}{\@spxthm{#1}{#2}{#3}{#4}}}
\def\@Thm#1#2#3{\topsep 1\p@ \@plus 1\p@ \@minus 1\p@
\@ifnextchar[{\@Ythm{#1}{#2}{#3}}{\@Xthm{#1}{#2}{#3}}}
\renewcommand\section{\@startsection{section}{1}{\z@}%
                       {-6\p@ \@plus -2\p@ \@minus -2\p@}%
                       {2\p@ \@plus 2\p@ \@minus 2\p@}%
                       {\normalfont\large\bfseries\boldmath
                        \rightskip=\z@ \@plus 8em\pretolerance=10000 }}
\renewcommand\subsection{\@startsection{subsection}{2}{\z@}%
                       {-6\p@ \@plus -2\p@ \@minus -2\p@}%
                       {2\p@ \@plus 2\p@ \@minus 2\p@}%
                       {\normalfont\normalsize\bfseries\boldmath
                        \rightskip=\z@ \@plus 8em\pretolerance=10000 }}
\renewcommand\subsubsection{\@startsection{subsubsection}{3}{\z@}%
                       {-9\p@ \@plus -3\p@ \@minus -3\p@}%
                       {-0.5em \@plus -0.22em \@minus -0.1em}%
                       {\normalfont\normalsize\bfseries\boldmath}}
\renewcommand\paragraph{\@startsection{paragraph}{4}{\z@}%
                       {-8\p@ \@plus -3\p@ \@minus -3\p@}%
                       {-0.5em \@plus -0.22em \@minus -0.1em}%
                       {\normalfont\normalsize\itshape}}
\def\@listI{\leftmargin\leftmargini \parsep \z@ \@plus 1\p@
  \topsep 3\p@ \@plus 1\p@ \@minus 1\p@ \partopsep 1\p@
  \itemsep 1\p@ \@plus .5\p@ \@minus .5\p@}
\let\@listi\@listI \@listi
\makeatother
\begin{document}

\maketitle

\begin{abstract}
\full{Global cryptocurrencies such as Bitcoin and Ether are unbacked and bear the cost of global consensus on every transaction.  }{}A grassroots coin is a claim on its issuer---any person, natural or legal\full{, including people, communities, cooperatives, corporations, banks, municipalities and governments}{}---who undertakes to accept their own coins for their offerings at the prices they post, and a transaction costs no more than operating a smartphone.  A grassroots coin is a unit of its issuer's debt, which the issuer must redeem at par (1-for-1) against any grassroots coin they hold.  Liquidity arises from mutual credit lines, formed by the voluntary swap of coins among persons who know and trust each other; coin redemption pegs mutually-liquid currencies at par and carries payments along chains of holdings.  

Here, we specify grassroots currencies by a grassroots social contract\full{ written in natural language (English)}{}, present its properties, and show two realisations of it: with pen-and-paper coins, and by formalising the contract and compiling it down an abstraction cascade to a working smartphone app.

The contract specifies four transactions---mint, swap, pay, redeem---and the coins in circulation form a directed multigraph, one edge per coin from its holder to its issuer, on which a redemption may act.  \full{The monetary properties of grassroots currencies are then properties of that graph.  }{}We prove that grassroots currencies are volitionally grassroots\full{---any group may operate on its own, and two groups become connected only by a swap willed by a member of each; adapt the classical liquidity ratios and creditworthiness measures to grassroots coins; and prove}{, and} that a community clears its debts into a designated member\full{, in any order and with no coordination,}{} exactly when no other member owes more grassroots coins than they hold, that the number of coins of that member a group can obtain by its own redemptions is the maximum flow from the group to that member in the graph of holdings, and that a run on a person is met in full exactly when they hold at least as many coins as they owe.

Grassroots currencies may thus allow local digital economies to form and grow without initial capital or external credit, harnessing mutual trust within communities into liquidity\full{, operating solely on the networked smartphones of their members, merging gradually into a global digital economy, and helping `banking the unbanked'}{}.
\end{abstract}

\section{Introduction}\label{sec:introduction}

\mypara{Grassroots currencies}
\looseness=-1 Global cryptocurrencies such as Bitcoin~\cite{nakamoto2008bitcoin} and Ether~\cite{buterin2014next} are unbacked and bear the cost of global consensus on every transaction.  A grassroots coin is a claim on its issuer---any person, natural or legal, including people, communities, cooperatives, corporations, banks, municipalities and governments---who undertakes to accept their own coins for their offerings at the prices they post, and a transaction costs no more than operating a smartphone.\iffull  Grassroots currencies are similar to fiat currencies in that scarcity is controlled by the issuer rather than by a protocol, in contrast to global cryptocurrencies such as Bitcoin~\cite{nakamoto2008bitcoin} or Ethereum~\cite{buterin2014next}.  They are similar to `inside money'~\cite{cavalcanti1999inside,cavalcanti1999model}---a medium of exchange backed by private credit---in that they are backed by claims on the issuer.\fi

\looseness=-1 Grassroots platforms are person-to-person smartphone apps that realise Proudhon's two-centuries-old notion of a social contract that can have multiple independent instances that may coalesce by mutual consent, possibly into a single global instance~\cite{proudhon2004general,shapiro2023grassrootsBA,shapiro2024grassroots}; grassroots currencies are one.  Grassroots currencies aim to provide a foundation for grassroots digital economies, allowing them to emerge without initial capital or external credit, operating solely on the networked smartphones of their members, and gradually merge into a global digital economy.  As such, grassroots currencies may, by harnessing mutual trust within communities, help `banking the unbanked'~\cite{bruhn2009economic,agarwal2017banking}.  These abilities may be most relevant where capital and external credit are scarce while trust within families and communities is not.

\mypara{Value and redemption}
\looseness=-1 Grassroots currencies are endowed with value by their issuer undertaking to redeem them against their offerings: Goods and services for people and corporations; interest and transaction fees for banks; and taxes for cities and states.  Critically, the offerings of each person include all the grassroots coins they hold, including coins issued by other persons.  A grassroots coin is a unit of its issuer's debt, which the issuer must redeem at par (1-for-1) against any grassroots coin they hold.

\looseness=-1 Coin redemption is key for a digital economy based on grassroots currencies to function, as it (\ia) makes coins issued by the same person fungible; (\ib) resolves double spending~\cite{lewis2023grassroots}; (\ic) allows credit to be settled and extinguished; (\id) allows chain payments across mutually-liquid currencies; and (\iie) prices mutually-liquid grassroots currencies at par, while their issuers meet presentations.

\mypara{Liquidity through mutual credit}
\looseness=-1 As each person prices their goods and services in terms of their own currency,  trade is impossible if no person holds another person's coins.  Liquidity arises from mutual credit lines, formed by the voluntary swap of coins among persons who know and trust each other; coin redemption pegs mutually-liquid currencies at par and carries payments along chains of holdings.  \iffull Such persons are family members, friends, peers, and colleagues; a community bank and its members; a corporation and its owners, employees, suppliers, and customers; philanthropic organisations, their donors, and the persons they support; a municipality and its local businesses.  \fi Liquidity is correlated with the amount and distribution of coins in circulation.  A grassroots coin in circulation is an outstanding obligation by its issuer to its holder, and the coins in circulation in a digital economy record the credit its members have extended one another.  The count itself is no measure across persons, each pricing their offerings in units of their own choosing. Since grassroots coins are redeemed at par, mutual credit is also at par: a credit line opened with a premium for one party can be closed by that party at once, keeping the premium.  \iffull Interest-bearing credit therefore needs coins with a maturity date---grassroots bonds, the subject of a separate paper~\cite{shapiro2026bonds}.\fi

\begin{figure}[t]
  \begin{center}
   \includegraphics[width=\full{\columnwidth}{6.5cm}]{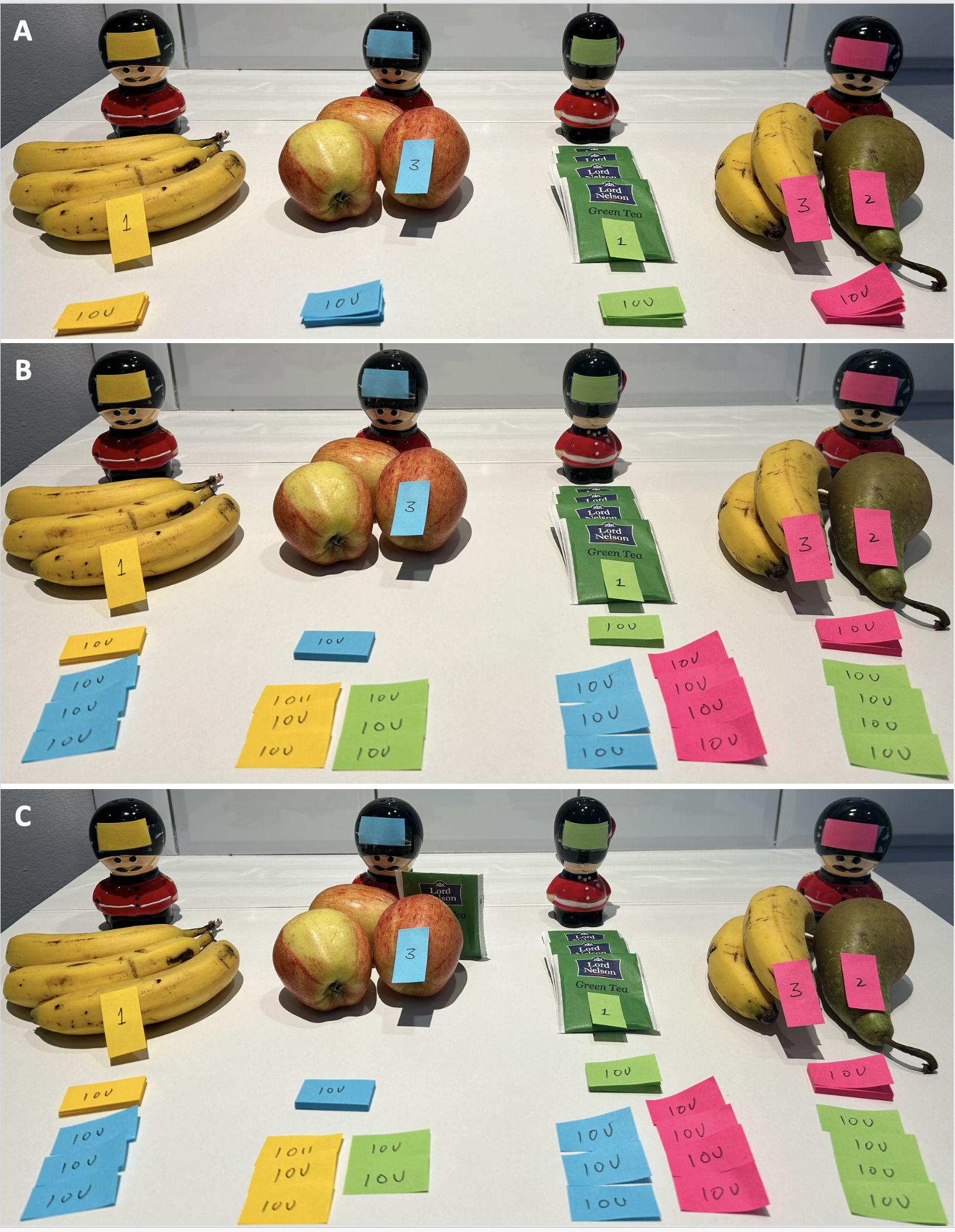}
  \end{center}
  \caption{Pen-and-paper grassroots currencies. \textmd{\textbf{A.}~Four persons, each with a pile of grassroots paper coins they have minted, each pricing their offerings in their own coins; no trade is possible, as no person holds another's coins.  \textbf{B.}~Mutual credit lines formed by voluntary swaps of grassroots paper coins among neighbours; trade can commence.  \textbf{C.}~The blue person has paid a green coin to the green person for a green tea bag.}}
  \label{figure:demo1}
\end{figure}

\mypara{Specification by a grassroots social contract}
\looseness=-1 Here, we specify grassroots currencies by a \emph{grassroots social contract}~\cite{shapiro2026formalising} written in natural language (English), present its properties (Section~\ref{sec:coins}), and show two realisations of it: with pen-and-paper coins (Figure~\ref{figure:demo1}, Section~\ref{sec:paper}), and by formalising the contract and compiling it down an abstraction cascade to a working smartphone app (Section~\ref{sec:contract}).  The clauses of the contract are what its parties undertake towards one another: minting, swap, payment, redemption, and pricing.  
\iffull A grassroots social contract is a social contract meeting Proudhon's conditions---reciprocal, imposing no obligation upon the parties except that which results from their personal promise, subject to no external authority, freely accepted and signed by all the participants---and two more, that people are free to deal with each other and that there is no external register of people.  A contract meeting these conditions has the property he envisioned: any two people may come into relation without anyone's leave, two communities that adopted the contract independently can later become one, and every participant in an act is a party~\cite{shapiro2026formalising}.  \fi

\looseness=-1 While the intended realisation of grassroots currencies is digital, the abstract contract does not dictate that.  Hence, we first illustrate the contract and its implications via a pen-and-paper realisation.  This simplifies their exposition and uncovers their inherent properties as a currency (Section~\ref{sec:paper}): a paper coin is a signed undertaking analogous to a promissory note, its obligations those of an ordinary contract, and a swap of coins at a meeting is physically atomic.

\mypara{From the contract to the app}
\looseness=-1 Down the abstraction cascade~\cite{shapiro2026volition} to the program every participant runs, the clauses become act schemas; the schemas compile to volition-guarded multiagent atomic transactions~\cite{lewis2026volitional}, each guarded by the persons whose consent it takes; and the transactions become the clauses of a program in vGLP~\cite{shapiro2026volition}, compiled to GLP~\cite{shapiro2025glp}, a multiagent concurrent logic programming language running on Dart for smartphone deployment.  The Grassroots Super-App~\cite{shapiro2026gsg} runs grassroots platforms as mini-apps, each a certified program it installs and activates and whose interface it renders, and grassroots currencies is one of them; a six-agent village market run on the compiled program exercises mutual credit, payments, swaps, redemption, and the sale of debt (Appendix~\ref{sec:implementation}).  The results of this paper are theorems about the transactions, so they hold under both realisations, on paper and digitally.

\mypara{From the currency to its graph}
\looseness=-1 The contract specifies four transactions---mint, swap, pay, redeem---and the coins in circulation form a directed multigraph, one edge per coin from its holder to its issuer, on which a redemption may act.  The monetary properties of grassroots currencies are then properties of that graph.  A coin an issuer holds of their own carries no value and no edge, so minting adds none; a swap of coins the two parties have issued adds an edge each way between them; a payment deletes an edge; and a redemption deletes the bearer's edge to the issuer and moves one of the issuer's edges to the bearer.  The out-degree of a person is the coins of others they hold, and their in-degree is the coins of theirs that others hold.  \iffull A community clears its debts into a designated member exactly when every other member's out-degree is at least their in-degree.  The coins of that member a group can obtain by its own redemptions are the maximum flow from the group to it.  A run on a person is met in full exactly when the coins they hold cover the coins they owe.  \fi The contract determines the transactions, the transactions determine the holdings, and these properties are read from the degrees and the maximum flows of the graph of holdings.

\mypara{Grassroots, solvency, liquidity, and runs on the bank}
\looseness=-1 We prove that grassroots currencies are volitionally grassroots---any group may operate on its own, and two groups become connected only by a swap willed by a member of each (Section~\ref{sec:contract}).  We adapt the classical liquidity ratios and creditworthiness measures to grassroots coins---the liquidity ratios of corporate finance and the creditworthiness measures of international monetary economics---introduce the community bank, and prove that a community clears its debts into a designated member, in any order and with no coordination, exactly when no other member owes more grassroots coins than they hold, that the number of coins of that member a group can obtain by its own redemptions is the maximum flow from the group to that member in the graph of holdings, and that a run on a person is met in full exactly when they hold at least as many coins as they owe---properties of the graph of holdings, needing no coordination among the persons (Section~\ref{sec:results}).  Grassroots currencies may thus allow local digital economies to form and grow without initial capital or external credit, harnessing mutual trust within communities into liquidity, operating solely on the networked smartphones of their members, merging gradually into a global digital economy, and helping `banking the unbanked'~\cite{bruhn2009economic,agarwal2017banking}.

\mypara{Paper structure}
\looseness=-1 Section~\ref{sec:coins} presents the grassroots currencies social contract and its properties.  Section~\ref{sec:paper} presents their pen-and-paper realisation.  Section~\ref{sec:contract} presents their digital realisation.  Section~\ref{sec:economics} presents the economics of grassroots currencies, assuming them broadly deployed.  Section~\ref{sec:related} places the work among related work, and Section~\ref{sec:conclusion} concludes.  The appendices carry \full{}{the pen-and-paper scenario (\ref{app:figures});  }act schemas of the contract (\ref{app:schemas});\full{}{  the abstraction cascade (\ref{app:cascade});}  program and the village market (\ref{sec:implementation}); the proofs (\ref{app:proofs}); more on liquidity (\ref{app:liquidity-more}); economic scenarios (\ref{sec:community-bank}); more on the digital realisation (\ref{app:digital-more}); implementation notes (\ref{app:impl-notes}); and related work, continued (\ref{sec:related-work}).

\section{Grassroots Currencies and Their Properties}\label{sec:coins}\label{sec:clauses}

Here we specify grassroots currencies by a social contract~\cite{shapiro2026formalising} and analyse their properties.

\subsection{The Grassroots Currencies Social Contract}

\looseness=-1 The grassroots currencies social contract~\cite{shapiro2026formalising} below describes what parties undertake towards one another.   A \emph{party} is a person, natural or legal (with the gender-neutral pronoun `they'), with \emph{Alice}, \emph{Bob} and \emph{Carol} as variables naming distinct parties to the contract.  A \emph{grassroots coin} is a unit of debt that names its issuer, with \emph{Alice-coin} referring to a grassroots coin issued by Alice.

\begin{eg}
\mypara{The Grassroots Currencies Social Contract}
\begin{enumerate}
\item \emph{Minting.}  Alice may decide to issue Alice-coins.
\item \emph{Swap.}  Alice and Bob may jointly decide to swap grassroots coins they hold. 
\item \emph{Payment.}  Alice may decide to pay Bob-coins they hold to Bob. 
\item \emph{Redemption.}  Alice may decide to redeem a Bob-coin they hold against any grassroots coin held by Bob.
\item \emph{Pricing.}  Alice prices their offerings in Alice-coins and accepts them in payment at those prices.
\end{enumerate}
\end{eg}

\subsection{The Transactions}\label{sec:gc-transactions-def}

\looseness=-1 We assume a potentially infinite set of persons $\Pi$ and consider only finite subsets of it; a grassroots $p$-coin, denoted \textcent$_{p}$, is a unit of debt issued by $p \in \Pi$, and \textcent$^k_{p}$ is a multiset of $k$ coins \textcent$_{p}$.  Each agent is a person operating a machine, and here the machine state of a party $p$ is their holdings $c_p$, a multiset of coins, initially empty.  A transaction changes the machine states of its participants and no one else's, is enabled when its precondition holds, and is \emph{guarded} by the persons among its participants whose consent it takes: it is taken only when every guard is willing.  Transactions that differ only in the configurations are \emph{equivalent}.  A run is \emph{safe} if every step is a transaction of the protocol.  We use $\cup$ and $\setminus$ for multiset sum and difference.  The four transactions are the following.

\begin{definition}[The Transactions of Grassroots Currencies]\label{def:gc-transactions}
\looseness=-1 For persons $p \ne q \in \Pi$, $u, v, r \in \Pi$ and amounts $k, k' \ge 1$, the \temph{transactions of grassroots currencies} are the volition-guarded transactions $(c \rightarrow c', Q')$ below, $Q'$ their guards~\cite{lewis2026volitional}, over the participants named, in which $c$ satisfies the precondition stated and $c'_x = c_x$ for every person $x$ not named:
\begin{enumerate}
    \item \temph{Mint$_p(k)$}, over $\{p\}$, guarded by $\{p\}$: $c'_p = c_p \cup \text{\textcent}^k_p$.
    \item \temph{Swap$_{p,q}(\text{\textcent}^k_u, \text{\textcent}^{k'}_v)$}, over $\{p,q\}$, guarded by $\{p,q\}$: precondition $\text{\textcent}^k_u \subseteq c_p$ and $\text{\textcent}^{k'}_v \subseteq c_q$; $c'_p = c_p \setminus \text{\textcent}^k_u \cup \text{\textcent}^{k'}_v$ and $c'_q = c_q \setminus \text{\textcent}^{k'}_v \cup \text{\textcent}^k_u$.
    \item \temph{Pay$_{p,q}(k)$}, over $\{p,q\}$, guarded by $\{p\}$: precondition $\text{\textcent}^k_q \subseteq c_p$; $c'_p = c_p \setminus \text{\textcent}^k_q$ and $c'_q = c_q \cup \text{\textcent}^k_q$.
    \item \temph{Redeem$_{p,q}(r)$}, over $\{p,q\}$, guarded by $\{p\}$: precondition $\text{\textcent}_q \in c_p$ and $\text{\textcent}_r \in c_q$; $c'_p = c_p \setminus \text{\textcent}_q \cup \text{\textcent}_r$ and $c'_q = c_q \setminus \text{\textcent}_r \cup \text{\textcent}_q$.
\end{enumerate}
Two transactions are \temph{equivalent} iff they have the same name and arguments, differing only in the configurations $c$ and $c'$.
\end{definition}

Mint, pay, and redeem are guarded by the initiator; a swap is guarded by both parties.  These transactions define a grassroots currencies protocol $\mathit{GC}$~\cite{lewis2026volitional}, with the following property.

\begin{lemma}[Conservation of Money, \cite{lewis2026volitional} for coins]\label{lem:conservation}
In any safe run $r$ of $\mathit{GC}$, the $p$-coins in any configuration $c\in r$ are exactly the $p$-coins minted by $p$ in the prefix of the run ending in $c$.
\end{lemma}

\subsection{The Graph of Holdings}\label{sec:liquidity}\label{sec:solvency}

\looseness=-1 The holdings of a grassroots economy form a directed multigraph, one edge per coin from its holder to its issuer, and that graph is the object of this section: the out-degree of a person is the coins of others they hold, their in-degree is the coins of theirs that others hold, and a redemption against a coin of another issuer shortens a path: where $p$ holds a $q$-coin and $q$ an $r$-coin, $r \ne q$, the two edges $p \rightarrow q \rightarrow r$ become the one edge $p \rightarrow r$, the $q$-coin returning to $q$, and where $r = p$ the two-cycle disappears by set-off.  Every such rewrite leaves the out-degree minus the in-degree of every person unchanged, and that invariant is what the results below rest on.  The persons act independently, each by their own rights, so we state and prove the answers in graph terms: the holdings form a graph, solvency depends on the graph degrees, and liquidity on the graph maximum flows.  No participant sees this graph; each sees their own holdings and what others disclose to them.  It is a mathematical construction for analysing the currency.

\subsection{Assets, Liabilities, and the Liquidity Ratio}\label{sec:ratio}

\looseness=-1 Let $\nu(p,q)$ be the number of $q$-coins held by $p$, where $p \ne q$ (zero if $p = q$, as $p$-coins held by $p$ carry no value).  The \emph{assets} of $p$ are the coins of other issuers they hold, $a_p := \sum_q \nu(p,q)$, and their \emph{liabilities} are their coins in circulation, $\ell_p := \sum_q \nu(q,p)$.  The \emph{liquidity ratio} of $p$ is $a_p / \ell_p$, undefined when $\ell_p = 0$.

A person is \emph{insolvent} when their liabilities exceed their assets, $a_p < \ell_p$: not every coin of theirs in circulation can be redeemed for a coin of another issuer, and no redemption of their own remedies that, as we prove below.

\subsection{Solvency, Liquidity, and Runs on the Bank}\label{sec:results}

\mypara{Computations}
\looseness=-1 A \emph{configuration} records, for every person, their holdings; a \emph{transition} leads from one configuration to the next---a transaction of the contract (Section~\ref{sec:clauses}) taking place.  A \emph{computation} is a sequence of configurations, each reached from the previous by a transition; it is \emph{safe} if every step is one, and a transaction is \emph{enabled} in a configuration if its preconditions hold there.  The possibility results below (Propositions~\ref{prop:graph-solvency} and~\ref{prop:graph-liquidity}) quantify over safe computations---what can happen; the liveness results (Propositions~\ref{prop:clearing-liveness} and~\ref{prop:run}) quantify over computations in which persons act on their rights---persons who can redeem do (Definition~\ref{def:maximal})---and state what must then happen.  In the volitional framework~\cite{lewis2026volitional}, that persons act on their rights is the liveness of correct runs (Section~\ref{sec:transactions}): an enabled class that is willed by its guards is eventually taken.  The proofs of the propositions below are in Appendix~\ref{app:proofs}.

\mypara{Chain redemption}  Chain redemption allows trade among parties that are connected in the holding graph, even if not adjacent.
\begin{proposition}[Chain Redemption]\label{prop:chain-redemption}
Let $p_0, p_1, \ldots, p_k$ be a sequence of distinct persons with $k \ge 1$ such that each $p_i$ ($0 \le i < k$) holds a $p_{i+1}$-coin.  Then from that configuration there is a sequence of $k-1$ redemptions by $p_0$ whose effect is that $p_0$ holds a $p_k$-coin.
\end{proposition}

\looseness=-1 A chain of holdings from $p$ to $q$---each person holding a coin of the next---is a chain of mutual liquidity: by Proposition~\ref{prop:chain-redemption}, a sequence of redemptions along the path leaves $p$ holding a $q$-coin.  Concurrent redemption chains that need the same coin interfere; what a group of persons can obtain by such chains is exactly the maximum-flow bound of Proposition~\ref{prop:graph-liquidity}.

\mypara{Graph solvency}
\looseness=-1 Can a community extinguish all its internal debts?  A debt is extinguished when the coin returns to its issuer, and a redemption does exactly that: the holder returns a coin to its issuer and takes in exchange any coin the issuer holds---as of right, the issuer being bound to redeem on presentation (Section~\ref{sec:coins}), without transferring value, and without changing any person's assets net of liabilities.  The propositions quantify over safe computations and do not assume a holder can inspect the debtor's holdings; in practice a debtor incapable of set-off discloses their holdings to the redeemer.  The results are stated relative to a designated person $s \in \Pi$, the \temph{sink}: the member with whom the community's net position is left---the community bank of \App{sec:bank}, or any member so designated.  Here we define the debts of a community as a graph, one edge per coin from holder to issuer, and prove: the community can extinguish all its debts into the sink by redemptions exactly when every member but the sink holds at least as many coins as they owe (Proposition~\ref{prop:graph-solvency}); and when it can, every maximal computation of clearing redemptions among its members ends with them cleared, in any order and with no coordination (Proposition~\ref{prop:clearing-liveness}).  Clearing uses Redeem only: a Pay transfers value and a swap requires the counterparty's consent.

\begin{definition}[Holding Graph, Sink-Closed]\label{def:holding-graph}
\looseness=-1 Given a configuration and a designated person $s \in \Pi$, the \temph{sink}, the \temph{holding graph} $G$ is the directed multigraph with the persons as vertices and an edge $u \rightarrow v$ for each $v$-coin that person $u \ne v$ holds.  A set of vertices containing $s$ is \temph{sink-closed} if every edge between two vertices other than $s$ has both endpoints in the set or neither.
\end{definition}

\looseness=-1 The out-degree and in-degree of $p$ in $G$ are their assets $a_p$ and liabilities $\ell_p$.  A sink-closed set of persons is a community whose only claims across its boundary are with the sink; the sink is a member of every such community.  A Redeem transaction in which the debtor is not $s$ and the coin taken is not issued by the debtor is a \emph{clearing redemption}.

\begin{definition}[Clearing into the Sink, Complete Clearing, Maximal]\label{def:clearing}\label{def:maximal}
\looseness=-1 Given a configuration and a sink-closed set $V$ of the holding graph, the holdings of $V$ \temph{clear into $s$} if some finite safe computation of clearing redemptions among members of $V$ leads to a configuration in which every coin of an issuer in $V$ other than $s$ is held by its issuer, and \temph{clear completely} if it leads to one in which every coin of an issuer in $V$ is held by its issuer.  A finite safe computation of clearing redemptions among members of $V$ is \temph{maximal} if no clearing redemption among members of $V$ is enabled in its last configuration.
\end{definition}

\begin{proposition}[Graph Solvency]\label{prop:graph-solvency}
The holdings of a sink-closed set $V$ clear into $s$ if and only if $a_p \ge \ell_p$ for every member $p$ of $V$ other than $s$, and cleared holdings leave each such member holding exactly $a_p - \ell_p$ coins of the sink.
\end{proposition}

\begin{proposition}[Clearing Liveness]\label{prop:clearing-liveness}
If $a_p \ge \ell_p$ for every member $p$ of $V$ other than $s$, then every maximal computation of clearing redemptions among members of $V$ ends with the holdings of $V$ cleared into $s$, after at most $\sum_{p \in V \setminus \{s\}} \ell_p$ redemptions.
\end{proposition}

\looseness=-1 Maximality assumes that members persistently will a clearing redemption while one is available; in a correct digital run an enabled transaction its guards will is eventually taken (Appendix~\ref{sec:implementation}), and whether a person wills it is theirs.  Complete clearing is the case of equality throughout\full{}{ (Corollary~\ref{cor:complete}, Appendix~\ref{app:proofs})}, and the special cases of Proposition~\ref{prop:graph-solvency} are in Appendix~\ref{app:liquidity-more}.

\iffull\begin{corollary}[Complete Clearing]\label{cor:complete}
The holdings of all persons clear completely if and only if $a_p = \ell_p$ for every person.
\end{corollary}

\begin{proof}
If $a_p = \ell_p$ for every person, then for any designated sink $s$ the holdings clear into $s$ by Proposition~\ref{prop:graph-solvency}, and each member other than $s$ then holds $a_p - \ell_p = 0$ coins of the sink, so no coin of $s$ is held by another and every coin is held by its issuer.  Conversely, in completely cleared holdings $a_p = \ell_p = 0$ for every person, and $a_p - \ell_p$ is invariant under Redeem (Appendix~\ref{app:proofs}), so $a_p = \ell_p$ in the given configuration.
\end{proof}
\fi

\mypara{Graph liquidity}
\looseness=-1 For clearing, any coin of the debtor may be taken; here a group needs specifically coins of the sink---$s$-coins.  How many can it obtain by its own redemptions?  Each redemption exchanges a coin the redeemer holds for a coin of the coin's issuer, so a member holding a coin whose issuer holds a coin whose issuer holds a coin, and so on to the sink, obtains an $s$-coin by chain redemption (Proposition~\ref{prop:chain-redemption}), consuming the coins along the chain.  Two chains that need the same coin cannot both complete, so the group obtains one $s$-coin per chain, over chains that share no coin; the maximum number of such chains is a maximum flow in the holding graph.  We recall the max-flow min-cut theorem~\cite{menger1927,ford1956maximal,elias1956note}: in a finite directed multigraph with unit edge capacities, the maximum flow from a set of source vertices to a sink vertex is realised by a maximum-size set of pairwise edge-disjoint directed paths from sources to the sink, and equals the minimum, over sets $S$ of vertices containing the sources and not the sink, of the number of edges from $S$ to its complement---a \emph{cut}.

\begin{definition}[Group Liquidity]\label{def:group-liquidity}
Given a configuration and a set $A$ of persons not containing $s$, the \temph{liquidity of $A$} is the maximum, over finite safe computations of Redeem transitions of members of $A$ from the configuration, of the number of $s$-coins members of $A$ hold in the last configuration.
\end{definition}

\begin{proposition}[Graph Liquidity]\label{prop:graph-liquidity}
The liquidity of $A$ is the maximum flow from $A$ to $s$ in $G$.
\end{proposition}

\looseness=-1 The edges here are coins already outstanding, not capacities to borrow or to route, and a member realises a path by exercising their own redemption rights one after another, with no intermediary asked to consent.  For a single person, $A = \{p\}$, the proposition states what they can obtain by themselves: their out-degree in $G$ counts the coins they hold---the numerator of their liquidity ratio---and the maximum flow is the number of $s$-coins they can reach by redemptions.  The proposition concerns unilateral capacity, what the members can obtain by their own rights, now: the transactions of persons not in $A$ are outside their control.  Group liquidity is subadditive, as paths of different members share edges.

\mypara{Runs on the bank}
In a run on the bank, the holders of a person's coins present them all at once, and the person meets each presentation with a coin they hold.  Each fulfilled presentation returns one $p$-coin to $p$ and takes one coin from $p$'s holdings, so a run is met in full under the same condition as solvency.

\begin{definition}[Run on the Bank]\label{def:run}
A \temph{run on} $p$ is a maximal finite safe computation of Redeem transactions, each surrendering a $p$-coin to $p$ and taking a coin of an issuer other than $p$; it is \temph{met in full} if $\ell_p = 0$ in its last configuration.
\end{definition}

\begin{proposition}[Run on the Bank]\label{prop:run}
Every run on $p$ is met in full if $a_p \ge \ell_p$, and otherwise ends with exactly $\ell_p - a_p$ $p$-coins in circulation, their presentations unfulfilled.
\end{proposition}

\section{Pen-and-Paper Realisation of Grassroots Currencies}\label{sec:paper}

Here we show how people can adopt and follow the contract with paper coins, in a scenario staged in Figure~\ref{figure:demo1} and continued in \App{app:figures}.

\subsection{Grassroots Paper Coins}\label{sec:paper-coins}

\looseness=-1 A person---natural or legal---is identified by their name and signature.  Persons include people, communities, cooperatives, corporations, banks, municipalities, and governments.  A pen-and-paper grassroots Alice-coin---a \emph{grassroots paper coin}---is written on paper and signed by Alice, in the form of a promissory note, stating the issuer, Alice, and two standing obligations:
\begin{enumerate}
    \item Alice prices their offerings in Alice-coins and accepts them in payment at these prices; and
    \item Alice prices every grassroots coin they hold, of any issuer, at par against Alice-coins.
\end{enumerate}
\looseness=-1 A grassroots paper coin is thus a signed, transferable undertaking of its issuer, analogous to a promissory note.  The obligations written on it are those of an ordinary contract.  It promises no sum certain in money, which the bills and notes acts require of a promissory note~\cite{boe1882,ucc3}, so whether it is one within those acts is a question for each jurisdiction.  Alice transfers any grassroots paper coin to Bob by signing it over to Bob and delivering it, so any two persons can transfer any grassroots paper coins between themselves, with no third party; the signatures on the coin record its provenance.  We assume that signatures cannot be forged and that signed papers cannot be duplicated (``double-spent'').  In a practical implementation, each Alice-coin will have a serial number and some value (here for simplicity they all are of identical and unit value) and be signed by Alice.

\mypara{The transactions with grassroots paper coins}
\looseness=-1 Alice and Bob swap coins (clause~2) at a meeting, by simultaneous signing-over and delivery, so the swap is physically atomic.  A payment and a redemption are Alice's rights under the written obligations: Bob must accept Bob-coins in payment (obligation~1) and must fulfil a presentation of a Bob-coin (obligation~2); declining either is a breach of the coin's terms, actionable at law.  A coin redeemed or paid to its issuer is self-held; whether the issuer keeps or cancels the paper is immaterial, as self-held coins carry no value (Section~\ref{sec:liquidity}).  

\iffull\iffull\subsection{A Scenario}\else\section{The Pen-and-Paper Demonstration}\fi\label{sec:scenario}\label{app:figures}

\mypara{Minting and pricing}
Figure~\ref{figure:demo1}A shows four persons---yellow, blue, green and red---each with a pile of grassroots paper coins they have minted, each pricing their offerings (bananas, apples, tea bags, and a pear) in terms of their own coins.  No trade can occur, as no person holds another person's coins.

\mypara{Mutual credit lines}
This is remedied by mutual credit lines, formed by the voluntary swap of grassroots coins among persons who know and trust each other.  Each of the four persons seems to trust their immediate neighbours, and therefore in Figure~\ref{figure:demo1}B neighbours establish mutual credit lines via coin swaps: four coins swapped between the green and red persons, and three coins between the green and blue and between the blue and yellow persons, resulting in a total of 20 grassroots coins in circulation.  Now trade can ensue.  For example, the blue person may buy a green tea bag from the green person using the green coin they hold, resulting in Figure~\ref{figure:demo1}C.

\mypara{Arbitrage}
In Figure~\ref{figure:demo1}C, the green person can buy a banana from the red person for 3 red coins, whereas the blue person can buy a banana from the yellow person for 1 yellow coin (Figure~\ref{figure:demo2}D) and then sell it to the green person for 2 blue coins, for a profit of 1 coin, with the green person saving 1 coin (Figure~\ref{figure:demo2}E).  Arbitrage trading can continue as long as prices differ and there is liquidity between the source and target of the arbitrage trade.

\begin{figure}[!htbp]
  \begin{center}
   \includegraphics[width=\columnwidth]{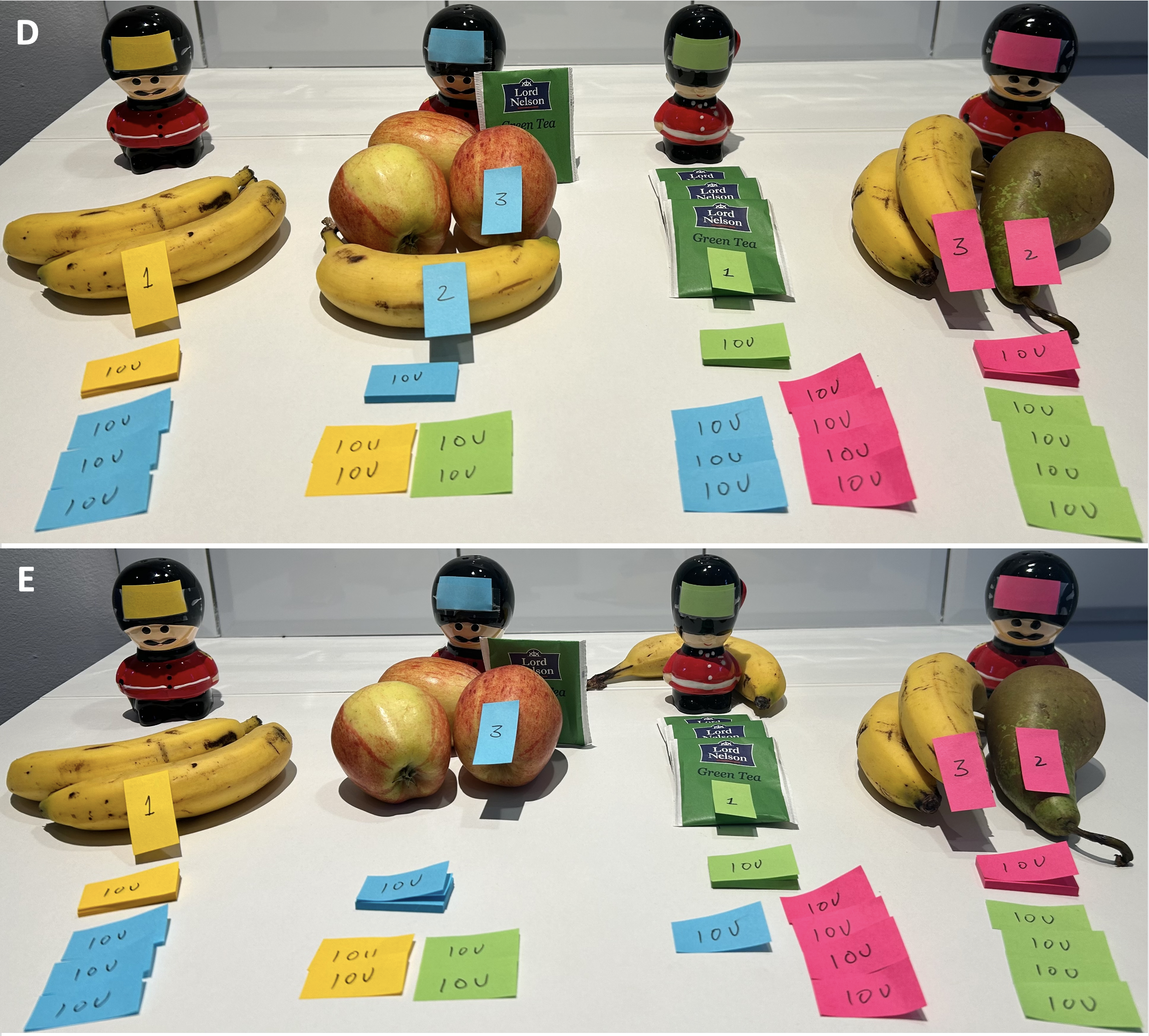}
  \end{center}
  \caption{Arbitrage. \textmd{In Figure~\ref{figure:demo1}C, the green person can buy a banana for 3 red coins from the red person, whereas the blue person can buy a banana from the yellow person for 1 yellow coin---an opportunity for arbitrage.  \textbf{D.}~The blue person has bought a banana from the yellow person for 1 yellow coin, and offers it for 2 blue coins.  \textbf{E.}~The green person has bought the banana from the blue person, saving 1 coin and making a 1-coin profit for the blue person.}}
  \label{figure:demo2}
\end{figure}

\begin{figure}[t]
  \begin{center}
   \includegraphics[width=\full{\columnwidth}{9cm}]{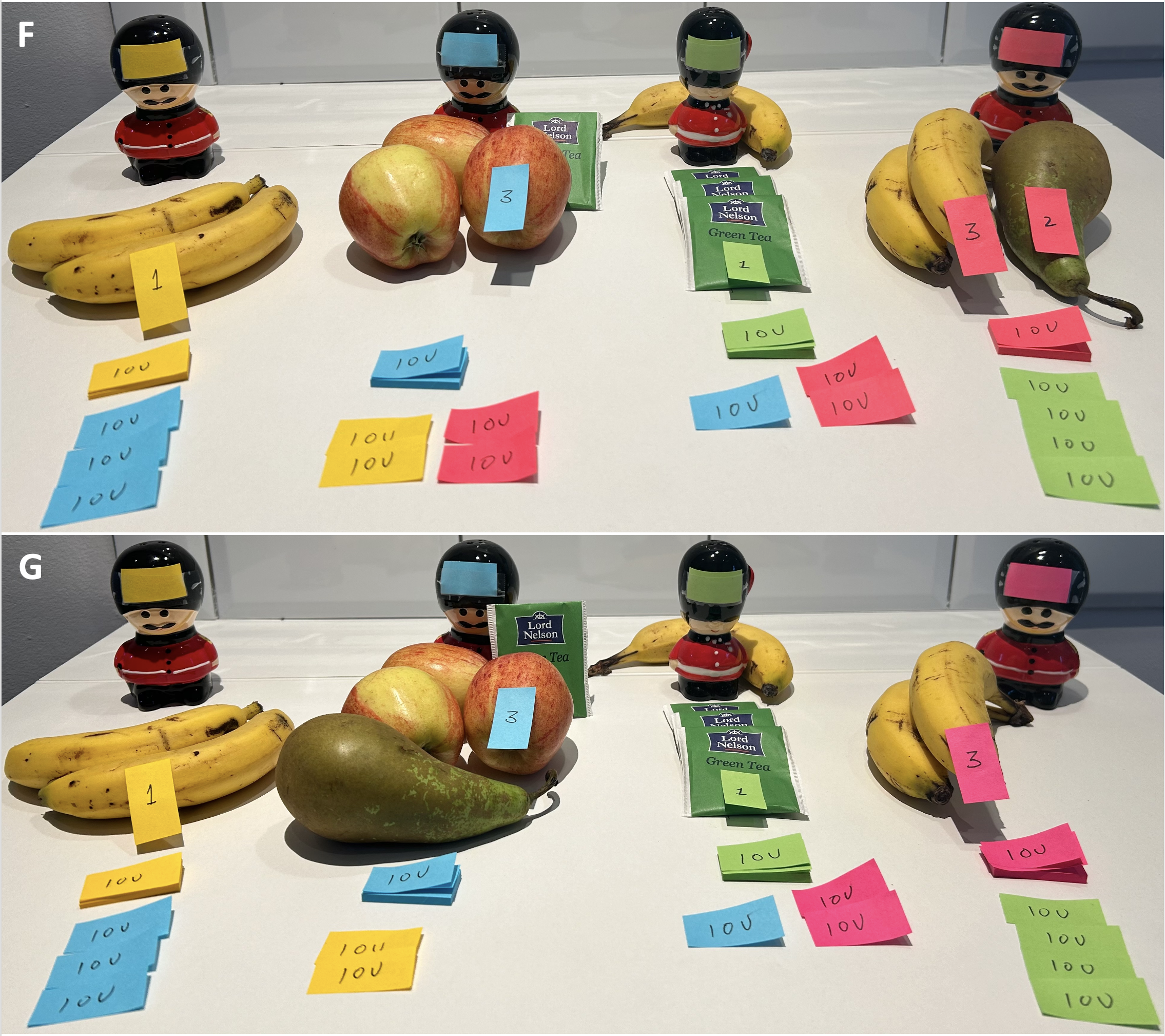}
  \end{center}
  \caption{Redemption and chain payment. \textmd{The blue person wishes to buy the pear of the red person, priced at two red coins, but holds none---only two green coins.  \textbf{F.}~The blue person redeems their two green coins from the green person, taking the two red coins the green person holds.  \textbf{G.}~They pay the two red coins to the red person for the pear.}}
  \label{figure:demo3}
\end{figure}

\mypara{Redemption and chain payment}
Having sold the banana to the green person (Figure~\ref{figure:demo2}), the blue person wishes to buy the pear of the red person, priced at two red coins, but holds none---only two green coins.  They therefore redeem their two green coins from the green person against the two red coins the green person holds (Figure~\ref{figure:demo3}F), and pay them to the red person for the pear (Figure~\ref{figure:demo3}G).  When Alice redeems a Bob-coin from Bob they may take an Alice-coin Bob holds, revoking credit Alice has given Bob; a fresh Bob-coin, to uncover a forgery; or a Carol-coin Bob holds, for a chain payment, for arbitrage, or to reduce their exposure to Bob and increase it to Carol.

\begin{figure}[t]
  \begin{center}
   \includegraphics[width=\full{\columnwidth}{9cm}]{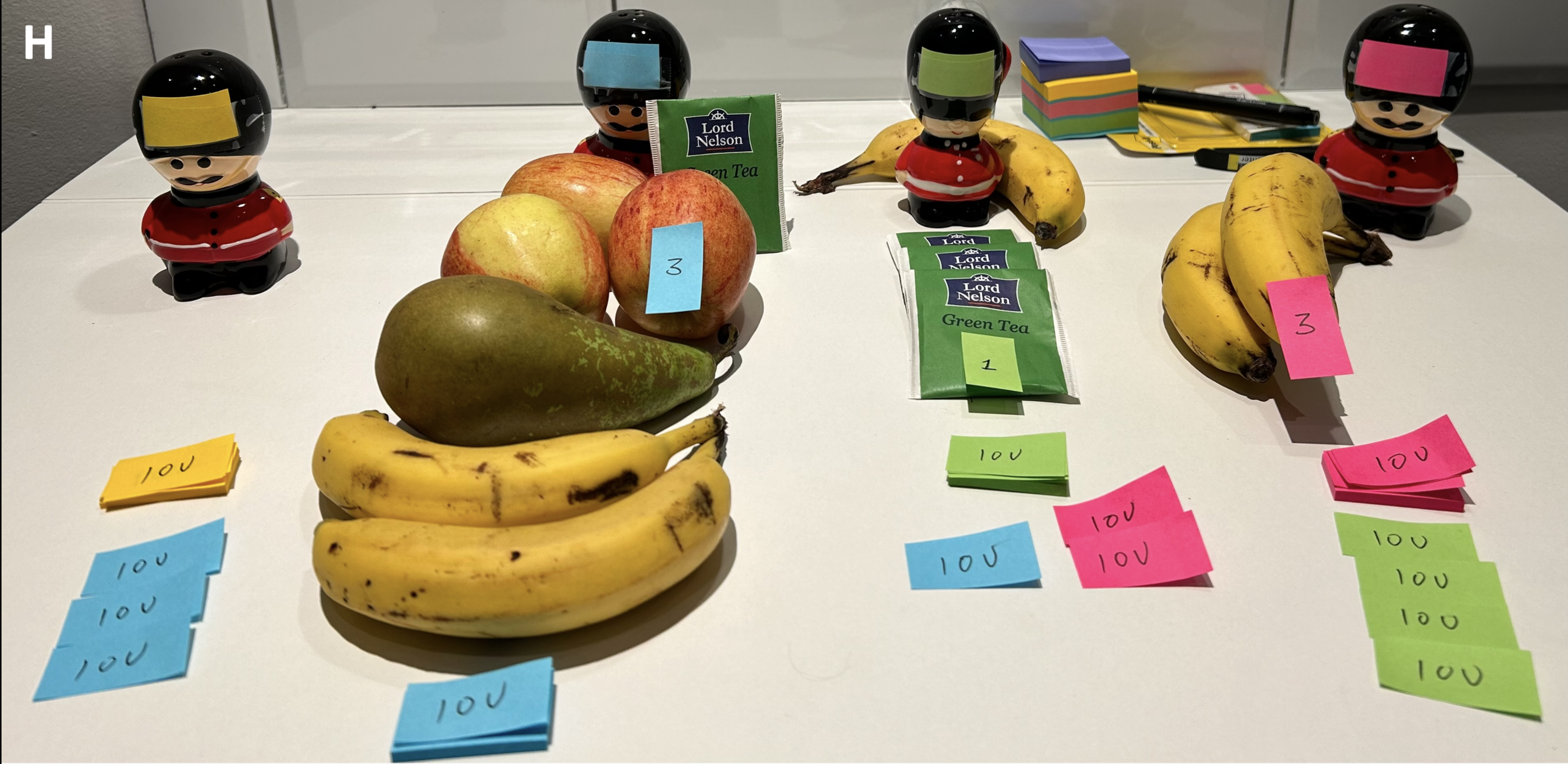}
  \end{center}
  \caption{Insolvency. \textmd{\textbf{H.}~After further trades, the blue person holds fruits and a tea bag but no coins of others, while four blue coins remain in circulation: their liabilities exceed their assets, and no presentation of a blue coin can be met with a coin of another issuer---the blue person is insolvent, though they may yet sell their goods to regain liquidity.}}
  \label{figure:demo4}
\end{figure}
\fi

\section{Digital Realisation of Grassroots Currencies}\label{sec:contract}

\looseness=-1 In the pen-and-paper realisation of Section~\ref{sec:paper}, people read the contract and follow it on signed notes.  Under it, safety and liveness are matters of law and of persons: violating an obligation of the contract is a breach, actionable at law, and so is declining a payment or a presentation (Appendix~\ref{sec:modalities}); and we may assume that whoever persistently wills a permitted transaction with a willing counterparty would eventually perform it.  In the digital realisation, both are properties of the program every participant runs~\cite{lewis2023grassroots}, and hold of a run in which every party runs it.  Here we derive that program from the contract.

\subsection{The Abstraction Cascade}\label{sec:cascade}

\iffull\iffull\else\section{The Abstraction Cascade}\label{app:cascade}\fi
\begin{figure}[!t]
\centering
\begin{tikzpicture}[
  layer/.style={draw=black!55, rounded corners=2pt, fill=gray!5,
                text width=\dimexpr\linewidth-1.1cm\relax, inner sep=4pt,
                align=left, font=\small},
  arr/.style={-{Latex[length=2mm,width=2.4mm]}, line width=1pt},
  lab/.style={font=\tiny, inner sep=1pt, midway, right=1mm}]
  \node[layer] (c1) {\textbf{(1) Grassroots social contract}~\cite{shapiro2026formalising}: the contract of Section~\ref{sec:clauses}.};
  \node[layer, below=4mm of c1] (c2) {\textbf{(2) Act schemas}~\cite{shapiro2026formalising}: the four schemas of Appendix~\ref{app:schemas}.};
  \node[layer, below=4mm of c2] (c3) {\textbf{(3) Volition-guarded multiagent atomic transactions}~\cite{lewis2026volitional}: Def.~\ref{def:gc-transactions}.};
  \node[layer, below=4mm of c3] (c4) {\textbf{(4) Communicating volitional agents}~\cite{keidar2026ivmat}: the currency agent, one per person, transacting over channels (Appendix~\ref{sec:derivation}).};
  \node[layer, below=4mm of c4] (c5) {\textbf{(5) Volition-guarded GLP}~\cite{shapiro2026volition}: \texttt{coins\_agent.vglp}, one volition-guarded clause per transaction (Appendix~\ref{sec:derivation}).};
  \node[layer, below=4mm of c5] (c6) {\textbf{(6) Grassroots Logic Programs}~\cite{shapiro2025glp,shapiro2026implementing,shapiro2026types}: \texttt{coins\_agent.glp}, emitted by the vGLP compiler with its mediator.};
  \node[layer, below=4mm of c6] (c7) {\textbf{(7) Dart and Flutter}~\cite{dart2024}: the GLP runtime for iPhone and Android.};
  \node[layer, below=4mm of c7] (c8) {\textbf{(8) The app}: the Grassroots Super-App~\cite{shapiro2026gsg}, running grassroots currencies as a mini-app, and the village market (Appendix~\ref{sec:village-demo}).};
  \draw[arr] (c1) -- node[lab] {[informal~\cite{shapiro2026formalising}]} (c2);
  \draw[arr] (c2) -- node[lab] {\cite{shapiro2026formalising}} (c3);
  \draw[arr] (c3) -- node[lab] {\cite{keidar2026ivmat}} (c4);
  \draw[arr] (c4) -- node[lab] {\cite{shapiro2026volition}} (c5);
  \draw[arr] (c5) -- node[lab] {\cite{shapiro2026volition}} (c6);
  \draw[arr] (c6) -- node[lab] {\cite{shapiro2026implementing}} (c7);
  \draw[arr] (c7) -- node[lab] {\cite{dart2024}} (c8);
\end{tikzpicture}
\caption{The abstraction cascade~\cite{shapiro2026volition} of grassroots currencies, from the contract to the app on each person's phone.  Each arrow but the first is an implementation of the abstraction above it by the one below, carrying the reference in which that implementation is defined, and implementations compose~\cite{shapiro2021multiagent}, so the app implements every abstraction above it; the first arrow is the realisation of the contract, writing its clauses as act schemas, and is the one step left to judgement.}
\label{fig:cascade}
\end{figure}
\fi

\looseness=-1 The derivation goes down the abstraction cascade~\cite{shapiro2026volition} of Figure~\ref{fig:cascade}\full{}{ (Appendix~\ref{app:cascade}), whose layers are defined in~\cite{shapiro2026formalising,lewis2026volitional,keidar2026ivmat,shapiro2026volition,shapiro2025glp,shapiro2026types,shapiro2026implementing,dart2024,shapiro2026gsg} and whose implementations compose~\cite{shapiro2021multiagent}}, from the contract in natural language (English) to the app on each person's phone, and every artefact of it is in this paper or in \repo.  The first two steps are defined in~\cite{shapiro2026formalising}.  Each clause that describes an act is written as an \emph{act schema}: who takes part in the act, whose decision it takes, and what each participant must hold, must not hold, gains and loses.  Each schema compiles to a \emph{volition-guarded transaction}~\cite{lewis2026volitional}: a change of the participants' holdings, carried out only when every participant whose decision the act takes wills it.  Three conditions on the schemas guarantee that the protocol is \emph{volitionally grassroots}: any group of parties can run it on its own, and two groups become connected only by a transaction that a member of each has decided.  The schemas of the contract, the modality of each clause---clauses 1--4 are enforced by the contract, clause 5 is undertaken---and the proof that the contract satisfies the three conditions are in Appendix~\ref{app:schemas}.  The transactions are those of Definition~\ref{def:gc-transactions}; the results of this paper are theorems about them, and the transactions of the pen-and-paper realisation are performed on notes, so the results hold under both realisations.

\mypara{One transaction through the cascade}
Figure~\ref{figure:swap-cascade} follows one transaction, the swap by which Alice and Bob open a mutual credit line, from the clause of the contract down the cascade.

\begin{figure}[t]
\centering
\scriptsize
\begin{tabular}{@{}p{0.07\textwidth}p{0.91\textwidth}@{}}
\textbf{(1)} & \emph{Swap.}  Alice and Bob may jointly decide to swap grassroots coins they hold. \\[2pt]
\textbf{(2)} & $\mathit{swap}(p?,q?;\,u,k,v,k') :$ at $p$, $-k\cdot\text{\textcent}(u),\ +k'\cdot\text{\textcent}(v)$; at $q$, $-k'\cdot\text{\textcent}(v),\ +k\cdot\text{\textcent}(u)$.  ``Alice and Bob may jointly decide'' is the two guarding roles of the swap; the mutual credit line is the binding sending $u$ to the party at $p$, $v$ to the party at $q$, and $k'$ to $k$. \\[2pt]
\textbf{(3)} & Swap$_{p,q}(\text{\textcent}^k_u, \text{\textcent}^{k'}_v)$, over $\{p,q\}$, guarded by $\{p,q\}$: precondition $\text{\textcent}^k_u \subseteq c_p$ and $\text{\textcent}^{k'}_v \subseteq c_q$; $c'_p = c_p \setminus \text{\textcent}^k_u \cup \text{\textcent}^{k'}_v$ and $c'_q = c_q \setminus \text{\textcent}^{k'}_v \cup \text{\textcent}^k_u$. \\[2pt]
\textbf{(4)} & The proposal reaches $q$ with the coins given and a response variable, and at $q$ it is the question of a responder, spawned per arriving proposal; the unbound response variable is the pending consent, and binding it commits or aborts the swap: the transaction is taken only when both guards are willing. \\[2pt]
\textbf{(5)} & At $p$, a request clause guarded by $p$, whose guard is the precondition $\text{\textcent}^k_u \subseteq c_p$, tested by selecting the coins given: \newline
{\ttfamily *(Friend, GiveCoin, GiveAmount, WantCoin, WantAmount)\newline
agent(Id, UserIn, PeerIn, Convs, Outs, Holdings, Serial) :-\newline
\hspace*{1.5em}ground(Id?), ground(Friend?), ground(GiveCoin?), integer(GiveAmount?), GiveAmount? >= 1, ground(WantCoin?), integer(WantAmount?), WantAmount? >= 1 |\newline
\hspace*{1.5em}select\_coins(GiveCoin?, GiveAmount?, Holdings?,\newline\hspace*{3em}Status, Give, Rest),\newline
\hspace*{1.5em}propose\_swap(Status?, Id?, Friend?, lot(GiveCoin?, GiveAmount?), lot(WantCoin?, WantAmount?), Give?, Rest?, UserIn?, PeerIn?, Convs?, Outs?, Serial?).}\newline
At $q$, the responder's one volition-guarded clause, the person's acceptance, with the decline as its else-branch: \newline
{\ttfamily *(yes, From?, Want?, Offered?)\newline
respond\_swap(offer(From, Want, Offered), Resp?, Ans?) :-\newline
\hspace*{1.5em}ground(From?), ground(Want?), ground(Offered?) |\newline
\hspace*{1.5em}accept\_answer(From?, Want?, Offered?, Resp, Ans)\newline
\hspace*{0.5em}*(no) decline\_answer(From?, Offered?, Resp, Ans).} \\[2pt]
\textbf{(6)} & The vGLP compiler produces \texttt{coins\_agent.glp}: the request clause becomes clauses that pose its question to the person and take its body on the answer, and the mediator shows the question as a compose form, with the counterparty, the coins given and the coins wanted; the clause of \texttt{respond\_swap} becomes one card per proposed swap, showing the proposer, the coins offered and the coins wanted, with the buttons Accept and Decline. \\[2pt]
\textbf{(7)} & The program runs on the GLP runtime, in Dart, on iPhone and Android. \\[2pt]
\textbf{(8)} & The Grassroots Super-App renders the form at the proposer and the card at the counterparty (Figure~\ref{fig:screens}); in the village market, Alice and Bob each mint 15~coins and swap them at par. \\
\end{tabular}
\caption{The swap, at each layer of the cascade of Figure~\ref{fig:cascade}, numbered as there.}
\label{figure:swap-cascade}
\end{figure}

\subsection{The Protocol}\label{sec:transactions}

\looseness=-1 The digital realisation begins by compiling the act schemas to volition-guarded multiagent atomic transactions~\cite{lewis2026volitional,shapiro2026formalising}, the transactions of Definition~\ref{def:gc-transactions}; from these the program every participant runs is derived (Appendix~\ref{sec:implementation}).  A run is \emph{live} if a class of equivalent transactions that is enabled and willed by its guards is eventually taken, and \emph{correct} if it is safe and live; a protocol is \emph{volitionally grassroots} if any group of persons may operate it on its own, and two groups become connected only by a transaction willed by a member of each~\cite{lewis2026volitional}.  In a correct run of a protocol, if an equivalence class of transactions is enabled and willed by its guards, then one of its members is eventually taken.

\mypara{Grassroots}
Coins among the members of a group require nothing outside the group, so any group may operate grassroots currencies on its own, and two groups may transact once their members meet and wish to do so.

\begin{corollary}[Volitionally Grassroots]\label{cor:gc-grassroots}
Grassroots currencies are volitionally grassroots.
\end{corollary}

\begin{proof}
\looseness=-1 The grassroots currencies social contract is syntactically grassroots at every value of its amounts (Appendix~\ref{app:schemas}), so the transactions of Definition~\ref{def:gc-transactions} are polite and the protocol over a polite set of volition-guarded transactions is volitionally grassroots~\cite{shapiro2026formalising} (Lemma~\ref{lem:amounts}).
\end{proof}

\subsection{From Volition-Guarded Transactions to a Smartphone App}\label{sec:digital}

\looseness=-1 The volition-guarded atomic transactions of Definition~\ref{def:gc-transactions} are implemented by communicating volitional agents~\cite{keidar2026ivmat}: the \emph{currency agent} every person runs, whose state is the person's holdings, transacting with the agents of other persons over channels.  The agent is a program in vGLP~\cite{shapiro2026volition}, GLP~\cite{shapiro2025glp} with volition guards, with one volition-guarded clause per transaction, guarded where the transaction is guarded; the vGLP compiler produces from it the GLP program every participant runs, together with the interface constructs it derives from the clauses.  The Grassroots Super-App~\cite{shapiro2026gsg} runs grassroots platforms as mini-apps, each a certified program it installs and activates and whose interface it renders; grassroots currencies is one of them.  The derivation, transaction by transaction, is in Appendix~\ref{sec:derivation}; the source and the compiled program are at \repo, in \texttt{programs/\allowbreak currencies/\allowbreak coins}.  A village scenario exercising the running code with all transactions by six people is shown in Appendix~\ref{sec:village-demo}.

\mypara{The issuer is the sole credit risk}
\looseness=-1 A grassroots coin is its issuer's debt, so trusting the issuer to honour it is unavoidable; we require that the issuer sign every transaction in their own coins, and may not withhold their signature from one the protocol permits, so that holdership of a coin is proven by the issuer's signatures, without trusting the holders it passed through.  This is the design of Grassroots Flash~\cite{lewis2023grassroots}; the argument, the treatment of double spending, and the issuer's signed log with its custodian, which we implement, are in Appendix~\ref{app:digital-more}.\full{}{  The program of Appendix~\ref{sec:implementation} is an executable realisation of the transactions of Definition~\ref{def:gc-transactions} among parties who run it, and the countersigned transfer of Appendix~\ref{app:digital-more} is a third level, in which the issuer signs every transfer of their own coins.  The third level is the secure coins of~\cite{eitan2026securing}: the transactions of Definition~\ref{def:gc-transactions} extended with the issuer's log of every transaction in their coins and with state custodians holding copies of it, the issuer a participant of every such transaction; that paper implements them by communicating volitional agents and proves that the implementation realises the specification at quiescence, conserves money, recovers the log exactly after the loss of a key or a machine, never double-spends after recovery, and finalises a requested payment while the issuer and a supermajority of its custodians stay live.}

\section{Grassroots Economics}\label{sec:economics}

This section assumes digital grassroots currencies broadly deployed and used, and presents the economics of the contract in that setting.

\subsection{The Contract in Economic Terms}\label{sec:properties}

\looseness=-1 In economic terms, a grassroots coin is an IOU---inside money~\cite{cavalcanti1999inside}---issued freely by its issuer, whose offerings and holdings back it.  Grassroots currencies rest on one principle: honour thy IOUs.  A person endows their coins with value by pricing their offerings in them (clause~5), and the offerings of each person include every coin they hold, of any issuer, priced at par (clause~4).  A swap between Alice and Bob in which Alice gives Bob Alice-coins and Bob gives Alice Bob-coins opens a mutual credit line between the two of them.

\mypara{Set-off and default}
\looseness=-1 Alice redeeming a Bob-coin against an Alice-coin that Bob holds is \emph{set-off}, where each gets their own coin back, at par.  An unfulfilled presentation to Alice of an Alice-coin is evidence for Alice's default, which in turn may cause a ``run on the bank'' on Alice and, if Alice turns out to be insolvent, the depreciation of the market value of Alice-coins.

\mypara{Backing}
\looseness=-1 The value of a coin rests on what its issuer accepts it for: the issuer undertakes to accept their own coins for their offerings at the prices they post, and to price every coin they hold at par against their own, so the value of a coin depends on the financial health of its issuer.  Coin redemption prices the coins of mutually-liquid persons at par while they meet presentations, hence at equal value.  The holder of an Alice-coin who loses trust in Alice may redeem it against any coin held by Alice, or sell it at a discount.

\iffull\mypara{Credit and the sale of debt as coin swaps}
Mutual credit, asymmetric credit, and the sale of debt are swaps of grassroots coins (clause~2), each between Alice and Bob and willed by both:
\begin{enumerate}
\item \textbf{Symmetric mutual credit.}  Alice and Bob each mint the same number of coins and swap them.  Each thereafter holds that many coins of the other, and trade between them can commence (Figure~\ref{figure:demo1}B).
\item \textbf{Asymmetric mutual credit.}  Alice, with high liquidity, charges a premium for their credit to Bob, a credit-less newcomer, giving fewer Alice-coins than the Bob-coins they receive.  The premium is Alice's profit once Bob has liquidity of their own, and Alice's loss if Bob spends the Alice-coins and disappears.  As coins are redeemable at par, Alice may redeem the Bob-coins they hold against Alice-coins at once and keep the premium; interest that is earned over time needs a maturity date, which is the subject of grassroots bonds~\cite{shapiro2026bonds}.
\item \textbf{Sale of debt.}  Alice sells Carol-coins to Bob at a discount, receiving fewer Bob-coins than the Carol-coins they give.  The debtor Carol is not a party; the discount reflects Alice's doubt about Carol's solvency.
\end{enumerate}

\mypara{Chain redemption}
Redemption reaches across chains of holdings, coin by coin: if each person of a sequence holds a coin of the next, the first can obtain a coin of the last by redeeming along the sequence.  Chain redemption enables long-range payments and arbitrage.  Consider a group of persons connected by such chains, rational and fully informed about the group.  A price difference between two of their coins is an arbitrage: the holder of the low-priced one chain-redeems it for the high-priced one and sells it at a profit, and such chain redemptions continue until the difference is gone or the liquidity along the path is exhausted---low-priced persons disconnecting from high-priced ones by exhausting the liquidity that connects them.

\mypara{Arbitrage}
Mutual liquidity allows arbitrage, staged in \App{app:figures}: while each person is free to price their offerings as they wish, a price difference for one good combined with liquidity allows arbitrage trading, with intermediaries splitting the profit, so the prices of that good stated in grassroots coins equalise within a liquid economy while the liquidity between the two sides persists, transaction frictions apart.
\fi

\subsection{Insolvency and Creditworthiness}\label{sec:insolvency}

\looseness=-1 The issuer may still hand back a coin of their own, which clause~4 permits and which leaves their liabilities where they were.  Solvency here is solvency at contractual par: the ratio prices every coin held at one, whatever its issuer's standing, and Appendix~\ref{app:liquidity-more} names the assets a lender discounts.  Default is sanctioned by the community~\cite{besley1995social}: holders redeem what the defaulter still holds, no one extends them new credit, they lose set-off and the ability to trade in their own currency, and the market value of their coins depreciates.  A solvent person may still lack the coins of a particular issuer; how many they can obtain by their own redemptions is a maximum flow in the graph of holdings (Proposition~\ref{prop:graph-liquidity}).  Figure~\ref{figure:demo4} stages insolvency with paper coins.

\mypara{Liquidity ratios}
\looseness=-1 The cash, quick, and current ratios of corporate finance stratify the assets by how soon they can be realised; every grassroots coin is redeemable now as of right, so for coins the three coincide in the liquidity ratio; the distinction they draw by asset class is drawn by maturity in grassroots bonds~\cite{shapiro2026bonds}.  A worked example on the village market and the dynamics of recovery from insolvency are in Appendix~\ref{app:liquidity-more}.

\mypara{Creditworthiness measures}
\looseness=-1 Determining creditworthiness is an art; still, grassroots currencies offer objective measures for it, analogous to those of international monetary economics: the net external position~\cite{imf2009bpm6} and velocity~\cite{fisher1911purchasing}.  The \emph{net external coin position} of $p$ is $\ell_p - a_p$, the number of $p$-coins held by others minus the number of coins of other issuers held by $p$; smaller is better, and its negation is the net investment position of $p$.  The \emph{change in net external coin position} of $p$ over a period is the difference between its values at the end and at the start of the period.  The \emph{velocity} of the currency of $p$ is the number of $p$-coins transferred during a period, divided by the average number of $p$-coins in circulation during it.  A lender assesses a borrower by these measures: a community bank opening credit lines with its members (\App{sec:bank}).  The positions sum to zero across the economy, so a member holding a large positive one holds coins others owe.

\iffull\subsection{The Community Bank}\label{sec:bank}

A community bank, or credit union, issues its own grassroots currency and opens credit lines with the members of its community at its discretion, swapping its coins for theirs; its members undertake to accept payments in the community currency, so transactions among liquid members may be one-step payments in it.  Such a payment is no transaction of Section~\ref{sec:clauses}: clause~3 pays a party coins of their own issue, and here the payee is given coins of the bank.  It is an \emph{extension} of the contract, by which a party undertakes to accept named currencies besides their own, and it is what a member's undertaking to the bank adds.  Without the extension a member reaches the seller's own currency by chain redemption, at the cost of the coins along the path.  The results of Section~\ref{sec:results} quantify over redemptions, so they hold of a configuration however it arose.  The bank needs no external funding to create the initial balances: the personal coins it holds back the community coins it has issued, and it manages its exposure to a member by redeeming that member's coins against coins of others the member holds, without withdrawing credit.  Backing is not capital: a bank whose assets equal its liabilities absorbs no loss.  The community mutual-credit scenario, the bank's credit lines with its members, the backing, and risk management by redemption are in \full{Appendix~\ref{sec:community-bank}}{\App{sec:community-bank} below}. The bank holds coins of every member and every member holds its coins, so a community's net position is naturally stated against it: whether the members can clear their debts among themselves by redemptions into it, leaving the balance with the bank; how many of the bank's coins a group of members can obtain by their own rights; and whether a run on the bank is met.  These are the results of Section~\ref{sec:results}, stated for a designated member---the bank, or any member so designated---the sink.
\fi

\section{Related Work}\label{sec:related}

The comparisons below are made in full in Appendix~\ref{sec:related-work}.

\mypara{Global cryptocurrencies and DeFi}
\looseness=-1 Bitcoin~\cite{nakamoto2008bitcoin} and Ethereum~\cite{buterin2014next,wood2014ethereum} keep one replicated ledger by consensus among miners or validators, so a transaction costs computing or locked capital and a fee for its place in the next block; grassroots currencies keep no shared ledger, and a transaction costs the operation of a networked device~\cite{lewis2023grassroots,shapiro2024grassroots}.  Smart contracts~\cite{szabo1997formalizing,buterin2014next,scilla2018} carry decentralised finance~\cite{werner2022sok,schar2021defi}: overcollateralised lending~\cite{gudgeon2020loanable,leshner2020compound}, automated market makers, and stablecoins.  Mutual credit and the sale of debt are swaps of coins (Section~\ref{sec:coins}), and the instruments of grassroots bonds are swaps of bonds and escrow agreements~\cite{shapiro2026bonds}, among persons who trust each other and without fees.  Flash loans~\cite{qin2021attacking}, automated market makers, and maximal extractable value~\cite{daian2020flash,qin2022quantifying} rest respectively on global atomicity, on pooled shared state, and on a global order of transactions, none of which the grassroots base protocol provides; 181 DeFi attack incidents are systematised in~\cite{zhou2023sokattacks}.

\iffull
\mypara{Payment and state channels}
\looseness=-1 Payment channels~\cite{poon2016bitcoin}, state channels~\cite{dziembowski2018general,coleman2018counterfactual}, and other layer-2 designs~\cite{gudgeon2020layer2} take transactions off the chain and return to it to settle a dispute; discreet log contracts~\cite{dryja2017discreet} are bilateral derivatives settled by an oracle's signature; cross-chain atomic swaps~\cite{herlihy2018atomic} exchange assets across two chains; and~\cite{boyen2016blockchain} replaces the chain by a DAG.  The abstract grassroots transaction is bilateral and settles with no chain: a party who disowns their signature is in default, remedied at law and made public by the record (Section~\ref{sec:insolvency}).
\fi

\mypara{Credit networks and personal currencies}
\looseness=-1 Mutual credit systems such as LETS keep one unit and one ledger: each member has an account with a central administration, a purchase runs the buyer's balance down or into debt, and the balances sum to zero~\cite{schraven2001mutual}, and the Swiss WIR, since 1934, is a bank keeping such accounts for its members, in credits denominated in but not redeemable for francs~\cite{stodder2009wir}; a grassroots coin is the debt of one person to another, in the issuer's own unit, with no administration.  Fugger's proposal of money as IOUs in a trust network~\cite{fugger2004money} is an early statement of the idea; the later XRP Ledger~\cite{schwartz2014ripple} realised IOU routing on a ledger with global consensus, and SilentWhispers~\cite{malavolta2017silentwhispers} without a ledger, for payment routing only.  Ghosh et al.\ define a trust network, in which a payment is routed as IOUs along links whose capacities are the credit each agent extends the next, and what a set of agents can pay one of them is the maximum flow from the set to it~\cite{ghosh2007trust}; credit networks~\cite{dandekar2011liquidity} route chain payments over bilateral credit lines, and Karlan et al.\ define network trust as a maximum flow of social collateral~\cite{karlan2009trust}; there the edges are capacities to borrow or to route, here they are coins outstanding, and the flow is what a group obtains by redemptions (Proposition~\ref{prop:graph-liquidity}).  Sardex~\cite{iosifidis2018sardex,fleischman2020liquidity} creates liquidity from trust within a community, with an operator, and Sarafu~\cite{mattsson2022sarafu} is one community currency administered by a foundation, its transactions duplicated onto a public blockchain; Holochain's white paper~\cite{harrisbraun2024holochain} describes a currency as accounting records on the chains of the two parties, each transfer an entry countersigned by spender and receiver, with no global consensus, in a unit the application fixes; Hypersyn~\cite{ramabaja2022hypersyn} is mutual credit with no ledger, consensus or validators, in which each node issues its own credit, an edge between two nodes holds reserves of both, and a payment moves the reserves by a constant-product rule, so nothing is held, transferred or redeemed; Circles~\cite{circles-UBI,circles2026docs} mints a personal currency for each account at a fixed rate, with demurrage, swapped one-for-one along a trust graph, and Trustlines~\cite{trustlines} hosts mutual-credit networks, both on blockchain infrastructure; FairTrade~\cite{hu2012fairtrade} has each peer of a file-sharing system issue a personal currency, spent by its issuer or by any peer who received it, accepted by its issuer for services and by other peers up to a credit limit each sets the issuer, with no central bank; personal cryptocurrencies were proposed in~\cite{delarouviere2013personal} and sold through an operator by Promify~\cite{promify2022}; and Worldcoin~\cite{worldcoin2022} is one global currency, at the antipode of a currency per person.  Prior work includes mutual credit, personal currencies, payments along paths of trust, routing without a ledger, agent-centric execution, and executable legal contracts.  Grassroots currencies are transferable person-issued debts redeemable as of right, specified by a social contract compiled to a program every participant runs, with no shared ledger and no global consensus, and with exact conditions on the holdings for clearing, liquidity and a run on the bank.

\iffull
\mypara{Clearing in financial networks}
\looseness=-1 Eisenberg and Noe~\cite{eisenberg2001systemic} prove that a clearing payment vector, every node paying its creditors pro rata, exists as a fixed point, the basis of the analysis of contagion~\cite{glasserman2016contagion}.  Grassroots clearing redeems unit coins in kind at the holder's initiative, and whether a community clears is a comparison of degrees, person by person (Proposition~\ref{prop:graph-solvency}).
\fi

\mypara{Community finance}
\looseness=-1 Group lending~\cite{morduch1999microfinance}; peer-to-peer lending, where friendship is a credit signal~\cite{lin2013friendship} and peers predict default better than credit scores~\cite{iyer2016screening}; rotating savings and credit associations~\cite{besley1993roscas,ardener1964roscas}; and hawala~\cite{elqorchi2003hawala} substitute trust within a community for institutional intermediation, as grassroots currencies do; community currencies, reviewed in~\cite{michel2015community}, contribute mostly to social sustainability, their economic benefits limited by their small scale; and 1.4 billion adults have no formal financial services~\cite{demirguckunt2022findex}.  The creditworthiness measures of Section~\ref{sec:insolvency} are those of international monetary economics, the net external position~\cite{imf2009bpm6} and velocity~\cite{fisher1911purchasing}; the scenarios of Appendix~\ref{sec:community-bank} draw on currency areas~\cite{mundell1961theory}, company scrip~\cite{green2011company}, and community basic income~\cite{assiagood,howitt2019roadmap}.

\iffull
\mypara{Bills, notes, and monetary theory}
\looseness=-1 Before banks dominated payment, debts circulated as bills of exchange and promissory notes~\cite{rogers1995early,santarosa2015financing}; a grassroots coin is a signed transferable IOU payable on demand, analogous to such a note but promising no sum certain in money, so its status under the bills and notes acts~\cite{boe1882,ucc3,geneva1930} and what its electronic signature establishes~\cite{eidas2014,esign2000,uncitral2001esignatures} are questions for each jurisdiction (Appendix~\ref{app:digital-more}).  Grassroots coins are inside money~\cite{cavalcanti1999inside,cavalcanti1999model}; money is memory~\cite{kocherlakota1998money}, and here the memory is the record each person keeps; a mutual credit line creates an asset and a liability at once, as bank lending creates deposits~\cite{mcleay2014money,palley2002endogenous}; and in the framework of~\cite{brunnermeier2019digitalization} grassroots currencies are peer-issued, record-based inside money with neither sovereign backing nor consensus.
\fi

\mypara{Specification and realisation}
\looseness=-1 Financial contracts have been specified declaratively since~\cite{peytonjones2000composing}, Marlowe~\cite{seijas2018marlowe} compiles such specifications to consensus-bound smart contracts, Scilla~\cite{scilla2018} is an intermediate-level smart-contract language, and the Accord Project~\cite{accord2026faq} writes legally binding agreements as text with machine-executable components, with no blockchain needed; here the currency is specified by a grassroots social contract~\cite{shapiro2026formalising} and realised as a digital social contract~\cite{cardelli2020digital} that the parties run, on any of the platform classes of~\cite{shapiro2025characterising}; on a grassroots platform a payment involves the payer, the payee and the issuer of the coins paid~\cite{lewis2023grassroots,shapiro2023gsn}.  Self-chosen identities cost nothing~\cite{douceur2002sybil}; a Sybil's coins are debt with no credit standing behind it, and a graph that friends can see exposes duplicate identities~\cite{shahaf2020genuine}.  Enforcement by mutual attestation, intended to leave an attested agent executing correctly or crashing rather than deviating arbitrarily~\cite{lamport1982byzantine}, is further work (Appendix~\ref{app:digital-more}).

\section{Conclusion}\label{sec:conclusion}

\looseness=-1 Grassroots currencies turn mutual trust into liquidity: a coin is its issuer's debt, a mutual credit line is a swap of coins between persons who trust each other, and coin redemption pegs the currencies of mutually-liquid persons at par and carries payments along chains of holdings.  The contract of Section~\ref{sec:clauses} states the currency, its four transactions enforced and its pricing undertaking witnessed by every coin in another's hands, and the coins in circulation form a graph on which a redemption acts locally.  Whether a community clears, what a group can obtain and whether a run is met are then read from the degrees and the maximum flows of that graph, and need neither coordination among the persons nor a shared ledger.  Each person acts by their own rights, and the properties of the whole economy follow from the graph of the holdings those transactions produce.

\looseness=-1 The contract has two realisations.  With pen and paper the obligations are written on the notes, and a breach of them is remedied under the law of the jurisdiction.  Digitally they are kept by every correct run of the program derived from the contract schema by schema, and the issuer signing every transaction in their own coins makes the issuer the sole credit risk---the risk already taken when their debt was accepted.  Holding a deployment to correct execution needs the assumptions of Appendix~\ref{app:digital-more}, with mutual attestation left to further work.

\looseness=-1 Grassroots bonds---coins with a maturity date, and with it the interest that redemption at par precludes, and the financial instruments built on them---are the subject of \full{a separate paper}{another paper}~\cite{shapiro2026bonds}.  Further work is the economy that is not a demonstration: issuers who default, and populations of thousands rather than six.  The secure coins---the issuer's log, its custodians, and recovery from the loss of a signing key or a device---are specified, implemented and proved in~\cite{eitan2026securing}.

\ifanon\else
\mypara{AI Disclosure}
\looseness=-1 The authors used Anthropic's Claude models to assist with two parts of this work.  First, the GLP code (Appendix~\ref{sec:implementation}) was derived from the contract of Section~\ref{sec:clauses} with substantial Claude assistance: the authors specified the contract, supplied GLP language documentation, iteratively reviewed the generated code against the specification and against a running implementation, and made all design decisions.  Second, the prose of the paper received Claude assistance for drafting and proofreading under the same review discipline.  The authors verified the correctness and originality of all content, including all references.  Responsibility for all content rests with the authors.
\fi

\bibliography{bib}

@misc{delarouviere2013personal,
  author = {Simon de la Rouviere},
  title = {In the Future, Everyone Will Have Their Own Cryptocurrency},
  year = {2013},
  howpublished = {\url{https://simondlr.tumblr.com/post/70089813484}},
  note 
  = {Retrieved 2022}
}

@misc{promify2022,
  author = {{Promify}},
  title = {Promify White Paper},
  year = {2022},
  howpublished = {\url{https://promify.io/whitepaper}},
  note = {Retrieved 2022}
}

@misc{worldcoin2022,
  author = {Alex Blania and Sam Altman and {the Worldcoin team}},
  title = {Worldcoin},
  year = {2022},
  howpublished = {\url{https://worldcoin.org}}
}

@misc{boe1882, author = {{United Kingdom}}, title = {Bills of Exchange Act 1882}, year = {1882}, howpublished = {45 \& 46 Vict.\ c.\ 61}}

@misc{ucc3, author = {{American Law Institute and Uniform Law Commission}}, title = {Uniform Commercial Code, Article 3: Negotiable Instruments}, year = {2002}}

@misc{geneva1930, author = {{League of Nations}}, title = {Convention Providing a Uniform Law for Bills of Exchange and Promissory Notes}, year = {1930}, howpublished = {Geneva, 7 June 1930}}

@book{rogers1995early, author = {Rogers, James Steven}, title = {The Early History of the Law of Bills and Notes}, publisher = {Cambridge University Press}, year = {1995}}

@article{santarosa2015financing, author = {Santarosa, Veronica Aoki}, title = {Financing Long-Distance Trade: The Joint Liability Rule and Bills of Exchange in Eighteenth-Century France}, journal = {The Journal of Economic History}, volume = {75}, number = {3}, pages = {690--719}, year = {2015}}

@misc{eidas2014,
  author = {{European Parliament and Council}},
  title  = {Regulation ({EU}) No 910/2014 on Electronic Identification and Trust Services for Electronic Transactions in the Internal Market ({eIDAS})},
  year   = {2014},
  note   = {Official Journal of the European Union, L 257}
}

@misc{esign2000,
  author = {{United States Congress}},
  title  = {Electronic Signatures in Global and National Commerce Act},
  year   = {2000},
  note   = {Public Law 106--229, 114 Stat.\ 464}
}

@misc{uncitral2001esignatures,
  author    = {{UNCITRAL}},
  title     = {Model Law on Electronic Signatures with Guide to Enactment},
  publisher = {United Nations},
  address   = {New York},
  year      = {2001}
}

@book{imf2009bpm6,
  author    = {{International Monetary Fund}},
  title     = {Balance of Payments and International Investment Position Manual},
  edition   = {Sixth},
  publisher = {International Monetary Fund},
  address   = {Washington, DC},
  year      = {2009}
}

@book{fisher1911purchasing,
  author    = {Irving Fisher},
  title     = {The Purchasing Power of Money: Its Determination and Relation to Credit, Interest and Crises},
  publisher = {Macmillan},
  address   = {New York},
  year      = {1911}
}

@article{eisenberg2001systemic,
  author  = {Larry Eisenberg and Thomas H. Noe},
  title   = {Systemic Risk in Financial Systems},
  journal = {Management Science},
  volume  = {47},
  number  = {2},
  pages   = {236--249},
  year    = {2001}
}

@article{glasserman2016contagion,
  author  = {Paul Glasserman and H. Peyton Young},
  title   = {Contagion in Financial Networks},
  journal = {Journal of Economic Literature},
  volume  = {54},
  number  = {3},
  pages   = {779--831},
  year    = {2016}
}

@article{menger1927,
  author  = {Karl Menger},
  title   = {Zur allgemeinen {K}urventheorie},
  journal = {Fundamenta Mathematicae},
  volume  = {10},
  pages   = {96--115},
  year    = {1927}
}

@article{ford1956maximal,
  author  = {Lester R. {Ford, Jr.} and Delbert R. Fulkerson},
  title   = {Maximal Flow through a Network},
  journal = {Canadian Journal of Mathematics},
  volume  = {8},
  pages   = {399--404},
  year    = {1956}
}

@article{elias1956note,
  author  = {Peter Elias and Amiel Feinstein and Claude E. Shannon},
  title   = {A Note on the Maximum Flow through a Network},
  journal = {IRE Transactions on Information Theory},
  volume  = {2},
  number  = {4},
  pages   = {117--119},
  year    = {1956}
}

@article{karlan2009trust,
  author  = {Karlan, Dean and Mobius, Markus and Rosenblat, Tanya and Szeidl, Adam},
  title   = {Trust and Social Collateral},
  journal = {The Quarterly Journal of Economics},
  volume  = {124},
  number  = {3},
  pages   = {1307--1361},
  year    = {2009}
}

@misc{shapiro2026bonds,
  author = {Ehud Shapiro},
  title  = {Grassroots Bonds as a Foundation for Market Liquidity},
  year   = {2026},
  howpublished = {arXiv:2603.13671},
  eprint = {2603.13671},
  archivePrefix = {arXiv},
  url    = {https://arxiv.org/abs/2603.13671}
}

@article{besley1995social,
  author  = {Timothy Besley and Stephen Coate},
  title   = {Group Lending, Repayment Incentives and Social Collateral},
  journal = {Journal of Development Economics},
  volume  = {46},
  number  = {1},
  pages   = {1--18},
  year    = {1995}
}

@misc{shapiro2026volition,
      title={Volition Elicitation:  Operational Semantics for People and Their Machines}, 
      author={Ehud Shapiro},
      year={2026},
      howpublished={arXiv:2607.14138},
      eprint={2607.14138},
      archivePrefix={arXiv},
      primaryClass={cs.DC},
      url={https://arxiv.org/abs/2607.14138}, 
}

@article{eitan2026securing,
      title={Securing People and their Machines Against Major Faults (Full Version)}, 
      author={Ohad Eitan and Idit Keidar and Ehud Shapiro},
  journal={Proceedings of The 28th International Symposium on Stabilization, Safety, and Security of Distributed Systems (SSS 2026), to appear},
  note={arXiv:2607.02304},
  year={2026},
  url = {https://arxiv.org/abs/2607.02304}
}

@article{lewis2026volitional,
  title={Volition-Guarded Multiagent Atomic Transactions: Describing People and their Machines},
  author={Lewis-Pye, Andy and Shapiro, Ehud},
  journal={arXiv preprint arXiv:2604.25596},
  year={2026},
  url = {https://arxiv.org/abs/2604.25596}
}

@misc{leshner2020compound,
  author       = {Leshner, Robert},
  title        = {Compound Governance},
  year         = {2020},
  howpublished = {Blog post, Compound Labs},
  url          = {https://medium.com/compound-finance/compound-governance-5531f524cf68}
}

@inproceedings{qin2021attacking,
  author    = {Qin, Kaihua and Zhou, Liyi and Livshits, Benjamin and Gervais, Arthur},
  title     = {Attacking the {DeFi} Ecosystem with Flash Loans for Fun and Profit},
  booktitle = {Financial Cryptography and Data Security -- 25th International Conference, {FC} 2021},
  year      = {2021},
  pages     = {3--32},
  publisher = {Springer},
  doi       = {10.1007/978-3-662-64322-8\_1}
}

@inproceedings{daian2020flash,
  author    = {Daian, Philip and Goldfeder, Steven and Kell, Tyler and Li, Yunqi and Zhao, Xueyuan and Bentov, Iddo and Breidenbach, Lorenz and Juels, Ari},
  title     = {Flash Boys 2.0: Frontrunning in Decentralized Exchanges, Miner Extractable Value, and Consensus Instability},
  booktitle = {2020 {IEEE} Symposium on Security and Privacy ({SP})},
  year      = {2020},
  pages     = {910--927},
  publisher = {IEEE},
  doi       = {10.1109/SP40000.2020.00040}
}

@inproceedings{qin2022quantifying,
  author    = {Qin, Kaihua and Zhou, Liyi and Gervais, Arthur},
  title     = {Quantifying Blockchain Extractable Value: How Dark is the Forest?},
  booktitle = {2022 {IEEE} Symposium on Security and Privacy ({SP})},
  year      = {2022},
  pages     = {198--214},
  publisher = {IEEE},
  doi       = {10.1109/SP46214.2022.9833734}
}

@article{shapiro2025characterising,
  title={Characterising Global Platforms: Centralised, Decentralised, Federated, and Grassroots},
  author={Shapiro, Ehud},
  journal={arXiv preprint arXiv:2511.03286},
  year={2025},
  url = {https://arxiv.org/abs/2511.03286}
}

@article{shapiro2026types,
  title={Types for Grassroots Logic Programs (Full version)},
  author={Shapiro, Ehud},
  journal={arXiv preprint arXiv:2601.17957},
  year={2026},
  url = {https://arxiv.org/abs/2601.17957}
}

@article{shapiro2026implementing,
  title={Implementing Grassroots Logic Programs with Multiagent Transition Systems and AI (Full Version)},
  author={Shapiro, Ehud},
  journal={arXiv:2602.06934.  Summary to appear in Proc. of LOPSTR+PPDP'26},
  year={2026},
  url = {https://arxiv.org/abs/2602.06934}
}

@misc{dart2024,
     author       = {{Google}},
     title        = {Dart Programming Language},
     year         = {2024},
  url = {https://dart.dev/}
}

@misc{nakamoto2008bitcoin,
  title={Bitcoin: A Peer-to-Peer Electronic Cash System},
  author={Nakamoto, Satoshi},
  year={2008},
  url={https://bitcoin.org/bitcoin.pdf}
}

@article{poon2016bitcoin,
  title={The bitcoin lightning network: Scalable off-chain instant payments},
  author={Poon, Joseph and Dryja, Thaddeus},
  journal={Draft version 0.5.9.2},
  year={2016},
  url={https://lightning.network/lightning-network-paper.pdf}
}

@article{schwartz2014ripple,
  author={Schwartz, David and Youngs, Noah and Britto, Arthur},
  title={The Ripple protocol consensus algorithm},
  journal={Ripple Labs White Paper},
  year={2014}
}

@misc{wood2014ethereum,
  title={Ethereum: A secure decentralised generalised transaction ledger},
  author={Wood, Gavin},
  year={2014}
}

@inproceedings{dziembowski2018general,
  title={General state channel networks},
  author={Dziembowski, Stefan and Eckey, Lisa and Faust, Sebastian and Malinowski, Daniel},
  booktitle={Proceedings of the 2018 ACM SIGSAC Conference on Computer and Communications Security},
  pages={949--966},
  year={2018}
}

@article{shapiro2025glp,
  title={GLP: A Grassroots, Multiagent, Concurrent, Logic Programming Language for AI (full version)},
  author={Shapiro, Ehud},
  journal={arXiv preprint arXiv:2510.15747, Summary in Proc. of ICLP'26, EPTCS 450, pp. 119--133},
  year={2025},
  url = {https://arxiv.org/abs/2510.15747}
}

@inproceedings{scilla2018,
  title={Scilla: a smart contract intermediate-level language},
  author={Sergey, Ilya and Kumar, Amrit and Hobor, Aquinas},
  booktitle={Proceedings of the 40th ACM SIGPLAN Conference on Programming Language Design and Implementation},
  pages={366--381},
  year={2019},
  organization={ACM}
}

@article{lamport1982byzantine,
  title={The Byzantine generals problem},
  author={Lamport, Leslie and Shostak, Robert and Pease, Marshall},
  journal={ACM Transactions on Programming Languages and Systems},
  volume={4},
  number={3},
  pages={382--401},
  year={1982},
  publisher={ACM}
}

@inproceedings{dandekar2011liquidity,
  title={Liquidity in credit networks: A little trust goes a long way},
  author={Dandekar, Pranav and Goel, Ashish and Govindan, Ramesh and Post, Ian},
  booktitle={Proceedings of the 12th ACM conference on Electronic commerce},
  pages={147--156},
  year={2011}
}

@misc{circles-UBI,
  title={Circles: A decentralised Universal Basic Income platform based on personal currencies},
  author={Circles},
  year={Retrieved 2021 https://joincircles.net/}
}

@misc{trustlines,
  title={Trustlines Network White Paper},
  author={Hees,Heiko and Friis, Gustav and Naerland, Kristoffer and Howitt, Aleeza and Kärki, Tatu and Fletcher, Andreas},
  year={2021, https://docs.trustlines.network/assets/pdf/Trustlines\_Network\_Whitepaper\_2021.pdf}
}

@misc{howitt2019roadmap,
  title={Roadmap to a Government-Independent Basic Income (UBI) Digital Currency},
  author={Howitt, A},
  year={2019}
}

@article{fugger2004money,
  title={Money as IOUs in social trust networks \& a proposal for a decentralized currency network protocol},
  author={Fugger, Ryan},
  journal={Hypertext document. Available electronically at http://ripple. sourceforge. net},
  volume={106},
  year={2004}
}

@article{mundell1961theory,
  title={A theory of optimum currency areas},
  author={Mundell, Robert A},
  journal={The American economic review},
  volume={51},
  number={4},
  pages={657--665},
  year={1961},
  publisher={JSTOR}
}

@book{green2011company,
  title={The company town: the industrial Edens and satanic mills that shaped the American economy},
  author={Green, Hardy},
  year={2011},
  publisher={ReadHowYouWant. com}
}

@article{kocherlakota1998money,
  title={Money is memory},
  author={Kocherlakota, Narayana R},
  journal={Journal of economic theory},
  volume={81},
  number={2},
  pages={232--251},
  year={1998},
  publisher={Elsevier}
}

@article{cavalcanti1999model,
  title={A model of private bank-note issue},
  author={Cavalcanti, Ricardo de O and Wallace, Neil},
  journal={Review of Economic Dynamics},
  volume={2},
  number={1},
  pages={104--136},
  year={1999},
  publisher={Elsevier}
}

@article{cavalcanti1999inside,
  title={Inside and outside money as alternative media of exchange},
  author={Cavalcanti, Ricardo de O and Wallace, Neil},
  journal={Journal of Money, Credit and Banking},
  pages={443--457},
  year={1999},
  publisher={JSTOR}
}

@article{bruhn2009economic,
  title={The economic impact of banking the unbanked: evidence from Mexico},
  author={Bruhn, Miriam and Love, Inessa},
  journal={World bank policy research working paper},
  number={4981},
  year={2009}
}

@article{agarwal2017banking,
  title={Banking the unbanked: What do 255 million new bank accounts reveal about financial access?},
  author={Agarwal, Sumit and Alok, Shashwat and Ghosh, Pulak and Ghosh, Soumya and Piskorski, Tomasz and Seru, Amit},
  journal={Columbia Business School Research Paper},
  number={17-12},
  year={2017}
}

@article{lewis2023grassroots,
  title={Grassroots Flash: A Payment System for Grassroots Cryptocurrencies},
  author={Lewis-Pye, Andrew and Naor, Oded and Shapiro, Ehud},
  journal={arXiv preprint arXiv:2309.13191},
  year={2023},
  url = {https://arxiv.org/abs/2309.13191}
}

@inproceedings{shapiro2023gsn, 
author={Shapiro, Ehud},
title={Grassroots Social Networking: Serverless, Permissionless Protocols for Twitter/LinkedIn/WhatsApp},
year = {2023},
isbn = {979-8-4007-0225-9/23/09},
publisher = {Association for Computing Machinery},
doi = {10.1145/3599696.3612898},
booktitle = {OASIS ’23},
location = {Rome, Italy},
}

@inproceedings{shapiro2023grassrootsBA, 
author={Shapiro, Ehud},
 title={Grassroots Distributed Systems: Concept, Examples, Implementation and Applications (Brief Announcement)
},year = {2023},
publisher = {LIPICS},
booktitle = {37th International Symposium on Distributed Computing (DISC 2023). (Extended version: arXiv:2301.04391)},
address = {Italy},
notes={Extended version: arXiv preprint arXiv:2301.04391},
pages = {47:1--47:7},
  url = {https://arxiv.org/abs/2301.04391}
}

@article{shapiro2024grassroots,
  title={A Grassroots Architecture to Supplant Global Digital Platforms by a Global Digital Democracy},
  author={Shapiro, Ehud},
  journal={arXiv:2404.13468, Proceedings of DAWO'24},
  year={2024},
  url = {https://arxiv.org/abs/2404.13468}
}

@article{shapiro2021multiagent,
  title={Multiagent Transition Systems for Composing Fault-Resilient Protocol Stacks},
  author={Shapiro, Ehud},
  journal={arXiv preprint arXiv:2112.13650},
  year={2021},
  url = {https://arxiv.org/abs/2112.13650}
}

@article{buterin2014next,
  title={A next-generation smart contract and decentralized application platform},
  author={Buterin, Vitalik},
  journal={white paper},
  volume={3},
  number={37},
  year={2014},
  url = {https://ethereum.org/whitepaper/}
}

@inproceedings{shahaf2020genuine,
  title={Genuine Personal Identifiers and Mutual Sureties for Sybil-Resilient Community Growth},
  author={Shahaf, Gal and Shapiro, Ehud and Talmon, Nimrod},
  booktitle={International Conference on Social Informatics},
  pages={320--332},
  year={2020},
  organization={Springer},
  publisher={Springer},
  address={EU},
  doi = {10.1007/978-3-030-60975-7_24}
}

@misc{cooperatives,
  title={Cooperative identity, values \& principles},
  url={https://www.ica.coop/en/cooperatives/cooperative-identity},
  author={ICA},
  year={Retrieved 2021}
}

@book{proudhon2004general,
  title={General idea of the revolution in the nineteenth century},
  author={Proudhon, Pierre-Joseph and Robinson, John Beverley},
  year={2004},
  publisher={Courier Corporation},
  note={Original work published 1851}
}

@article{assiagood,
  title={Good Dollar Experiment},
  author={Assia, Yoni and Ross, Omri},
  year={2021, https://www.gooddollar.org/wp-content/uploads/2018/11/GD-Wealth-Distribution-Position-Paper.pdf}
}

@inproceedings{douceur2002sybil,
  title={The sybil attack},
  publisher={Springer},
  address={EU},
  author={Douceur, John R},
  booktitle={Proceedings of the international workshop on peer-to-peer systems},
  pages={251--260},
  year={2002},
}

@misc{proof,
  title={Proof of Stake FAQ, Ethereum Wiki},
  author={Nakamoto, Satoshi and Bitcoin, A},
  howpublished = { \\ \MYhref{https://eth.wiki/en/concepts/proof-of-stake-faqs}{https://eth.wiki/en/concepts/proof-of-stake-faqs}},
  year={2019}
}

@inproceedings{cardelli2020digital,
  title={Digital social contracts: A foundation for an egalitarian and just digital society},
  author={Cardelli, Luca and Orgad, Liav and Shahaf, Gal and Shapiro, Ehud and Talmon, Nimrod},
  booktitle={CEUR Proceedings of the First International Forum on Digital and Democracy},
  volume={2781},
  pages={51--60},
  year={2020},
  organization={CEUR-WS},
  url = {https://ceur-ws.org/Vol-2781/paper5.pdf}
}

@inproceedings{herlihy2018atomic,
  title={Atomic cross-chain swaps},
  author={Herlihy, Maurice},
  booktitle={Proceedings of the 2018 ACM symposium on principles of distributed computing},
  pages={245--254},
  year={2018}
}

@misc{szabo1997formalizing,
  title={Formalizing and Securing Relationships on Public Networks},
  author={Szabo, Nick},
  journal={First Monday},
  volume={2},
  number={9},
  year={1997}
}

@inproceedings{werner2022sok,
  author    = {Werner, Sam M. and Perez, Daniel and Gudgeon, Lewis and Klages-Mundt, Ariah and Harz, Dominik and Knottenbelt, William J.},
  title     = {{SoK}: Decentralized Finance ({DeFi})},
  booktitle = {Proceedings of the 4th ACM Conference on Advances in Financial Technologies (AFT '22)},
  pages     = {446--461},
  year      = {2022},
  publisher = {ACM},
  doi       = {10.1145/3558535.3559780}
}

@article{schar2021defi,
  author  = {Sch\"{a}r, Fabian},
  title   = {Decentralized Finance: On Blockchain- and Smart Contract-Based Financial Markets},
  journal = {Federal Reserve Bank of St.\ Louis Review},
  volume  = {103},
  number  = {2},
  pages   = {153--174},
  year    = {2021},
  doi     = {10.20955/r.103.153-74}
}

@inproceedings{gudgeon2020loanable,
  author    = {Gudgeon, Lewis and Werner, Sam M. and Perez, Daniel and Knottenbelt, William J.},
  title     = {{DeFi} Protocols for Loanable Funds: Interest Rates, Liquidity and Market Efficiency},
  booktitle = {Proceedings of the 2nd ACM Conference on Advances in Financial Technologies (AFT '20)},
  pages     = {92--112},
  year      = {2020},
  publisher = {ACM},
  doi       = {10.1145/3419614.3423254}
}

@inproceedings{zhou2023sokattacks,
  author    = {Zhou, Liyi and Xiong, Xihan and Ernstberger, Jens and Chaliasos, Stefanos and Wang, Zhipeng and Wang, Ye and Qin, Kaihua and Wattenhofer, Roger and Song, Dawn and Gervais, Arthur},
  title     = {{SoK}: Decentralized Finance ({DeFi}) Attacks},
  booktitle = {2023 IEEE Symposium on Security and Privacy (S\&P)},
  year      = {2023},
  publisher = {IEEE},
  doi       = {10.1109/SP46215.2023.10179435}
}

@inproceedings{peytonjones2000composing,
  author    = {Peyton Jones, Simon and Eber, Jean-Marc and Seward, Julian},
  title     = {Composing Contracts: An Adventure in Financial Engineering (Functional Pearl)},
  booktitle = {Proceedings of the 5th ACM SIGPLAN International Conference on Functional Programming (ICFP '00)},
  pages     = {280--292},
  year      = {2000},
  publisher = {ACM},
  doi       = {10.1145/351240.351267}
}

@inproceedings{seijas2018marlowe,
  author    = {Lamela Seijas, Pablo and Thompson, Simon},
  title     = {Marlowe: Financial Contracts on Blockchain},
  booktitle = {Leveraging Applications of Formal Methods, Verification and Validation (ISoLA 2018)},
  series    = {LNCS},
  volume    = {11247},
  pages     = {356--375},
  year      = {2018},
  publisher = {Springer},
  doi       = {10.1007/978-3-030-03427-6\_27}
}

@article{mattsson2022sarafu,
  author  = {Mattsson, Carolina E. S. and Criscione, Teodoro and Ruddick, William O.},
  title   = {Sarafu Community Inclusion Currency 2020--2021},
  journal = {Scientific Data},
  volume  = {9},
  pages   = {426},
  year    = {2022},
  doi     = {10.1038/s41597-022-01539-4}
}

@article{iosifidis2018sardex,
  author  = {Iosifidis, George and Charette, Yanick and Airoldi, Edoardo and Littera, Giuseppe and Tassiulas, Leandros and Christakis, Nicholas},
  title   = {Cyclic Motifs in the {Sardex} Monetary Network},
  journal = {Nature Human Behaviour},
  volume  = {2},
  pages   = {822--829},
  year    = {2018},
  doi     = {10.1038/s41562-018-0450-0}
}

@inproceedings{gudgeon2020layer2,
  author    = {Gudgeon, Lewis and Moreno-Sanchez, Pedro and Roos, Stefanie and McCorry, Patrick and Gervais, Arthur},
  title     = {{SoK}: Layer-Two Blockchain Protocols},
  booktitle = {Financial Cryptography and Data Security (FC 2020)},
  series    = {LNCS},
  volume    = {12059},
  pages     = {201--226},
  year      = {2020},
  publisher = {Springer},
  doi       = {10.1007/978-3-030-51280-4\_12}
}

@article{mcleay2014money,
  author  = {McLeay, Michael and Radia, Amar and Thomas, Ryland},
  title   = {Money Creation in the Modern Economy},
  journal = {Bank of England Quarterly Bulletin},
  volume  = {Q1},
  pages   = {14--27},
  year    = {2014}
}

@techreport{brunnermeier2019digitalization,
  author      = {Brunnermeier, Markus K. and James, Harold and Landau, Jean-Pierre},
  title       = {The Digitalization of Money},
  institution = {National Bureau of Economic Research},
  type        = {Working Paper},
  number      = {26300},
  year        = {2019},
  doi         = {10.3386/w26300}
}

@inproceedings{malavolta2017silentwhispers,
  author    = {Malavolta, Giulio and Moreno-Sanchez, Pedro and Kate, Aniket and Maffei, Matteo},
  title     = {{SilentWhispers}: Enforcing Security and Privacy in Decentralized Credit Networks},
  booktitle = {Proceedings of the Network and Distributed System Security Symposium (NDSS 2017)},
  year      = {2017},
  publisher = {Internet Society}
}

@misc{boyen2016blockchain,
  author       = {Boyen, Xavier and Carr, Christopher and Haines, Thomas},
  title        = {Blockchain-Free Cryptocurrencies: A Framework for Truly Decentralised Fast Transactions},
  year         = {2016},
  howpublished = {Cryptology ePrint Archive, Report 2016/871},
  note         = {Also published as ``Graphchain'' at BCC@AsiaCCS 2018}
}

@article{palley2002endogenous,
  author  = {Palley, Thomas I.},
  title   = {Endogenous Money: What It Is and Why It Matters},
  journal = {Metroeconomica},
  volume  = {53},
  number  = {2},
  pages   = {152--180},
  year    = {2002}
}

@article{morduch1999microfinance,
  author  = {Morduch, Jonathan},
  title   = {The Microfinance Promise},
  journal = {Journal of Economic Literature},
  volume  = {37},
  number  = {4},
  pages   = {1569--1614},
  year    = {1999},
  doi     = {10.1257/jel.37.4.1569}
}

@article{lin2013friendship,
  author  = {Lin, Mingfeng and Prabhala, Nagpurnanand R. and Viswanathan, Siva},
  title   = {Judging Borrowers by the Company They Keep: Friendship Networks and Information Asymmetry in Online Peer-to-Peer Lending},
  journal = {Management Science},
  volume  = {59},
  number  = {1},
  pages   = {17--35},
  year    = {2013},
  doi     = {10.1287/mnsc.1120.1560}
}

@article{iyer2016screening,
  author  = {Iyer, Rajkamal and Khwaja, Asim Ijaz and Luttmer, Erzo F. P. and Shue, Kelly},
  title   = {Screening Peers Softly: Inferring the Quality of Small Borrowers},
  journal = {Management Science},
  volume  = {62},
  number  = {6},
  pages   = {1554--1577},
  year    = {2016},
  doi     = {10.1287/mnsc.2015.2181}
}

@article{besley1993roscas,
  author  = {Besley, Timothy and Coate, Stephen and Loury, Glenn},
  title   = {The Economics of Rotating Savings and Credit Associations},
  journal = {American Economic Review},
  volume  = {83},
  number  = {4},
  pages   = {792--810},
  year    = {1993}
}

@article{ardener1964roscas,
  author  = {Ardener, Shirley},
  title   = {The Comparative Study of Rotating Credit Associations},
  journal = {The Journal of the Royal Anthropological Institute of Great Britain and Ireland},
  volume  = {94},
  number  = {2},
  pages   = {201--229},
  year    = {1964},
  doi     = {10.2307/2844382}
}

@book{elqorchi2003hawala,
  author    = {El~Qorchi, Mohammed and Maimbo, Samuel Munzele and Wilson, John F.},
  title     = {Informal Funds Transfer Systems: An Analysis of the Informal Hawala System},
  publisher = {International Monetary Fund},
  series    = {IMF Occasional Paper},
  number    = {222},
  year      = {2003},
  isbn      = {978-1-58906-183-5}
}

@book{demirguckunt2022findex,
  author    = {Demirg\"u\c{c}-Kunt, Asli and Klapper, Leora and Singer, Dorothe and Ansar, Saniya},
  title     = {The Global Findex Database 2021: Financial Inclusion, Digital Payments, and Resilience in the Age of {COVID-19}},
  publisher = {World Bank},
  year      = {2022},
  doi       = {10.1596/978-1-4648-1897-4}
}

@misc{shapiro2026gsg,
  author = {Ohad Eitan and Idit Keidar and Ehud Shapiro},
  title  = {The Grassroots Social Graph: A Secure Operating System for Multiagent, Distributed, Peer-to-Peer, Smartphone-Based Platforms},
  year   = {2026},
  note   = {In preparation}
}

@misc{dryja2017discreet,
  author = {Dryja, Thaddeus},
  title  = {Discreet Log Contracts},
  year   = {2017},
  note   = {MIT Digital Currency Initiative},
  url    = {https://adiabat.github.io/dlc.pdf}
}

@misc{coleman2018counterfactual,
  author = {Coleman, Jeff and Horne, Liam and Xuanji, Li},
  title  = {Counterfactual: Generalized State Channels},
  year   = {2018},
  note   = {Whitepaper},
  url    = {https://github.com/counterfactual/paper}
}

@article{shapiro2026formalising,
  author  = {James Golike and Andy Lewis-Pye and Ehud Shapiro},
  title   = {Formalising and Implementing Grassroots Social Contracts: From Legal Text to Grassroots Platforms (Full Version)},
  year    = {2026},
  journal = {arXiv:2609.06231}
}

@unpublished{keidar2026ivmat,
  author = {Idit Keidar and Ehud Shapiro and friends},
  title  = {Fault-Resilient Implementations of Volition-Guarded Multiagent Atomic Transactions},
  note   = {In preparation},
  year   = {2026}
}

@article{schraven2001mutual,
  title={Mutual Credit Systems and the Commons Problem: Why Community Currency Systems such as {LETS} Need Not Collapse Under Opportunistic Behaviour},
  author={Schraven, Jorim},
  journal={International Journal of Community Currency Research},
  volume={5},
  number={4},
  year={2001},
  doi={10.15133/j.ijccr.2001.002}
}

@techreport{harrisbraun2024holochain,
  title={Holochain: Distributed Coordination by Scaled Consent, not Global Consensus},
  author={Harris-Braun, Eric and Brock, Arthur and d'Aoust, Paul},
  institution={Holochain Foundation},
  type={White paper},
  number={v2.0},
  year={2024},
  note={\url{https://www.holochain.org/documents/holochain-white-paper-2.0.pdf}}
}

@misc{circles2026docs,
  title={Welcome to {Circles}: {Circles} Documentation},
  author={{Circles}},
  howpublished={\url{https://docs.aboutcircles.com/}},
  year={2026},
  note={Accessed 21 September 2026}
}

@misc{accord2026faq,
  title={{FAQ}: {Accord Project} Frequently Asked Questions},
  author={{Accord Project}},
  howpublished={\url{https://docs.accordproject.org/docs/accordproject-faq/}},
  year={2026},
  note={Accessed 21 September 2026}
}

@inproceedings{ghosh2007trust,
  title={Mechanism Design on Trust Networks},
  author={Ghosh, Arpita and Mahdian, Mohammad and Reeves, Daniel M. and Pennock, David M. and Fugger, Ryan},
  booktitle={Internet and Network Economics, Third International Workshop (WINE 2007)},
  series={Lecture Notes in Computer Science},
  volume={4858},
  year={2007},
  publisher={Springer}
}

@article{stodder2009wir,
  title={Complementary credit networks and macroeconomic stability: {Switzerland}'s {Wirtschaftsring}},
  author={Stodder, James},
  journal={Journal of Economic Behavior \& Organization},
  volume={72},
  number={1},
  pages={79--95},
  year={2009},
  doi={10.1016/j.jebo.2009.06.002}
}

@article{michel2015community,
  title={Community currencies and sustainable development: A systematic review},
  author={Michel, Arnaud and Hudon, Marek},
  journal={Ecological Economics},
  volume={116},
  pages={160--171},
  year={2015},
  doi={10.1016/j.ecolecon.2015.04.023}
}

@misc{ramabaja2022hypersyn,
  title={Hypersyn: A Peer-to-Peer System for Mutual Credit},
  author={Ramabaja, Lum},
  howpublished={arXiv:2206.04049},
  year={2022}
}

@article{hu2012fairtrade,
  title={Peer-to-peer indirect reciprocity via personal currency},
  author={Hu, Yi and Bhuyan, Laxmi N. and Feng, Min},
  journal={Journal of Parallel and Distributed Computing},
  volume={72},
  pages={1045--1054},
  year={2012},
  doi={10.1016/j.jpdc.2012.04.008}
}

@article{fleischman2020liquidity,
  title={Liquidity-Saving through Obligation-Clearing and Mutual Credit: An Effective Monetary Innovation for {SMEs} in Times of Crisis},
  author={Fleischman, Toma{\v{z}} and Dini, Paolo and Littera, Giuseppe},
  journal={Journal of Risk and Financial Management},
  volume={13},
  number={12},
  pages={295},
  year={2020},
  doi={10.3390/jrfm13120295}
}

\newpage
\appendix
\iffull\else\fi
\section{The Act Schemas of the Contract}\label{app:schemas}

This appendix writes the clauses of Section~\ref{sec:clauses} as act schemas~\cite{shapiro2026formalising}, states the modality of each clause, and shows the contract syntactically grassroots.  We extend the schemas of~\cite{shapiro2026formalising}, whose atoms name people, with numeric variables, for amounts, and with multiplicities on the atoms added and deleted: at each value of its numeric variables a schema is a schema of~\cite{shapiro2026formalising} with an atom repeated as its multiplicity says, so at each value of the numeric variables the four schemas are a contract of~\cite{shapiro2026formalising}, and the instances taken together are a countable set of schemas, whose classes over any finite set of persons are countable~\cite{lewis2026volitional}.  A contract of~\cite{shapiro2026formalising} is a finite set of act schemas, so its theorem is not applied to that countable set; Lemma~\ref{lem:amounts} establishes the conclusion for it through politeness, a condition on transactions and not on schemas.  Compiling a schema~\cite{shapiro2026formalising} determines, at each binding of its variables, a transaction whose participants are the roles and whose guards are the guarding roles, enabled when the roles hold what they require, with effect what they add and delete; two transactions are equivalent if they are one schema at one binding.  The four schemas below compile to the transactions of Definition~\ref{def:gc-transactions}.

\subsection{The Act Schemas}\label{sec:gc-schemas}

One predicate, \textcent, of arity one: \textcent$(u)$ in the state of $p$ is a coin issued by $u$ and held by $p$.  The roles are $p$ and $q$, Alice and Bob of Section~\ref{sec:clauses}; the party variables are $u$, $v$ and $r$; the numeric variables are $k$ and $k'$, for amounts.  The four clauses that describe acts become four schemas.

\begin{center}
\footnotesize
\setlength{\tabcolsep}{3pt}
\begin{tabular}{lll}
\hline
\emph{schema} & \emph{at its first role} & \emph{at its second role} \\
\hline
$\mathit{mint}(p?;\,k) :$ & $+k\cdot\text{\textcent}(p)$ & \\
$\mathit{swap}(p?,q?;\,u,k,v,k') :$ & $-k\cdot\text{\textcent}(u),\ +k'\cdot\text{\textcent}(v)$ & $-k'\cdot\text{\textcent}(v),\ +k\cdot\text{\textcent}(u)$ \\
$\mathit{pay}(p?,q;\,k) :$ & $-k\cdot\text{\textcent}(q)$ & $+k\cdot\text{\textcent}(q)$ \\
$\mathit{redeem}(p?,q;\,r) :$ & $-\text{\textcent}(q),\ +\text{\textcent}(r)$ & $-\text{\textcent}(r),\ +\text{\textcent}(q)$ \\
\hline
\end{tabular}
\end{center}

\noindent ``Alice may decide'' is the single guarding role of mint, of pay and of redeem; ``Alice and Bob may jointly decide'' is the two guarding roles of the swap.  The amount minted is the argument $k$ of mint.  ``Bob-coins they hold'' and ``a Bob-coin they hold'' are that pay and redeem require of $p$ a coin whose argument is $q$; ``any coin held by Bob'' is redeem's party variable $r$, unconstrained.  The mutual credit line of clause~2 is the swap at a binding sending $u$ to the party at $p$, $v$ to the party at $q$, and $k'$ to $k$; asymmetric credit is the same binding with $k' < k$, and the sale of debt a binding sending $u$ to a third party (Section~\ref{sec:coins}).

\subsection{The Modality of Each Clause}\label{sec:modalities}

The definitions are~\cite{shapiro2026formalising}'s.  A \emph{clause} of a contract is a proposition about the conduct of its parties, kept or breached by that conduct, and a party it names as answerable is its \emph{obligor}; the runs of the protocol realising the contract \emph{determine} the clause if whether it is kept is a function of the run, and a transaction \emph{may breach} it if it occurs in some correct run that breaches it.  A transaction is \emph{witnessed against} a party if after it another of its participants holds an atom naming that party.  A clause with obligor $o$ is \emph{enforced} if the runs determine it and every correct run keeps it; \emph{attested} if some transaction may breach it and every transaction that may is witnessed against $o$; and \emph{undertaken} if no transaction may breach it, the runs do not determine it, and some transaction is witnessed against $o$ or $o$'s undertaking is held by another party.

Clauses 1--4 are enforced.  Each permits an act and obliges nothing beyond it, so the runs determine it and every correct run keeps it.  Two of them oblige a counterparty, and the schemas above make the obligation self-executing: ``Alice may decide'' in clause~3 is that pay guards at the payer only and adds the coins at the payee, and in clause~4 that redeem guards at the bearer only and deletes the taken coin at the issuer.  Neither the payee nor the issuer is asked.  No clause of the contract is attested: no act of it carries a warranty, and pay adds a coin naming its issuer and not its payer, so no clause against the payer could be~\cite{shapiro2026formalising}.

Clause~5 is undertaken.  Selling at an advertised price is no act of the contract, so no transaction of it may breach the clause and no run determines whether it is kept; and it is witnessed against the issuer: the swap opening a mutual credit line adds \textcent$(p)$ in the state of the party at $q$, naming the issuer $p$, so the coin in another party's holding is the record of the undertaking.  The second obligation written on the paper coin (Section~\ref{sec:paper-coins})---the standing offer at par---is not a further undertaking: its exercise is the redemption act, enforced as clause~4.

\subsection{The Contract Is Grassroots}\label{sec:gc-grassroots}

A contract is syntactically grassroots if it has an unobstructed introductory act, every predicate its acts require has traceable provenance, and every act not willed by all its parties runs only between parties who already record each other; the protocol realising a syntactically grassroots contract is volitionally grassroots~\cite{shapiro2026formalising}.  The three conditions quantify over the people a state records, and the numeric arguments of the schemas above name no person, so the conditions apply to them as stated; the conclusion for the countable set is Lemma~\ref{lem:amounts}.

\mypara{Introduction}  The swap at the binding of the mutual credit line is an introductory act: it is guarded in both roles, it adds \textcent$(q)$ at $p$, naming $q$, and \textcent$(p)$ at $q$, naming $p$.  It is unobstructed: what each role requires, $k$ coins of its own issue, is added by mint, a schema of arity one, guarded in its role, requiring and forbidding nothing; and no role forbids anything.

\mypara{Provenance}  \textcent\ has traceable provenance.  Mint adds \textcent$(p)$ at $p$, naming its own role.  Swap adds \textcent$(v)$ at $p$ and \textcent$(u)$ at $q$, and $v$ and $u$ are arguments of the atoms the other role requires.  Pay adds \textcent$(q)$ at $q$, naming a role.  Redeem adds \textcent$(r)$ at $p$, where $r$ is an argument of the atom $q$ requires, and \textcent$(q)$ at $q$, naming a role.

\mypara{Volition}  Of the schemas of arity two, swap is guarded in both roles.  Pay and redeem are guarded at $p$ only, and in each $p$ requires \textcent$(q)$, an atom of traceable provenance naming $q$, so the role graph is connected: a party pays or presents a coin only to an issuer whose coin they already hold, which is to say already records.

\medskip
\noindent The contract is therefore syntactically grassroots at every value of its numeric variables.

\begin{lemma}[Amounts]\label{lem:amounts}
The transactions of Definition~\ref{def:gc-transactions} are a polite set~\cite{shapiro2026formalising}, and the protocol over them is volitionally grassroots.
\end{lemma}

\begin{proof}
A set of volition-guarded transactions is \emph{polite} if it is open---any two persons can always be brought to a configuration at which some transaction coupling them is enabled---and \emph{closed}---every transaction enabled at the end of a safe run has all its participants in one instance of that run, or a guard of it in every instance holding one of them; the protocol over a polite set is volitionally grassroots~\cite{shapiro2026formalising}.  Both are conditions on the set of transactions, and neither depends on the number of schemas.

The checks above are per schema and per binding and hold at every value of the numeric variables, so each instance is a contract of~\cite{shapiro2026formalising} that is syntactically grassroots.  Closure quantifies over the runs of the protocol over the whole set, and a run of the union is no run of a single instance, so it is proved here of the union.  \textcent{} has traceable provenance in every schema of every instance, and the induction on a run by which~\cite{shapiro2026formalising} proves that no transition puts an edge whose ends lie in different instances uses only the schema inducing the current transition, so it carries to the union unchanged: whenever $p$ holds a $q$-coin, $p$ and $q$ lie in one instance of the run.  Mint is unary; Swap is guarded in both its roles; and a Pay or a Redeem enabled between $p$ and $q$ requires $p$ to hold a $q$-coin, so the two already lie in one instance.  The union is therefore closed.  For openness take any two persons $p \ne q$: each mints one coin of their own by Mint$_p(1)$ and Mint$_q(1)$, unary transactions, and Swap$_{p,q}(\text{\textcent}_p, \text{\textcent}_q)$ is then enabled, and it is a transaction of the union.  The union is therefore polite, and the protocol over it is volitionally grassroots.
\end{proof}

\noindent Any group of persons therefore operates grassroots currencies on its own, and two groups become connected only by a swap willed by a member of each (Corollary~\ref{cor:gc-grassroots}, Section~\ref{sec:transactions}).

\iffull\else\fi
\section{From the Contract to the App}\label{sec:implementation}\label{app:derivation}

The digital realisation of Section~\ref{sec:contract} in full: the program every participant runs, derived from the transactions of Definition~\ref{def:gc-transactions} one by one down the cascade of Figure~\ref{fig:cascade}, and the village market run on it.  The Grassroots Super-App~\cite{shapiro2026gsg} runs the compiled program as a mini-app.  Under the digital realisation some of the legal obligations of the coins become self-executing.

\mypara{Volitions and guards}
In the pen-and-paper scenario, a person consents by taking part in the transaction; digitally, standing consents are recorded.  The assumption of Section~\ref{sec:results}---that persons act on their rights---becomes the correctness requirement of the framework: in a correct run, a transaction that is enabled and willed by its guards is eventually taken~\cite{lewis2026volitional}.  The protocol so specified is \emph{volitionally grassroots}: two groups become coupled only by a first transaction willed by a member of each (Corollary~\ref{cor:gc-grassroots}).

\mypara{Three levels, and what this paper establishes of each}
The transactions of Definition~\ref{def:gc-transactions} are the abstract acts, each over the two persons the act names, and the results of Sections~\ref{sec:results} and~\ref{sec:contract} are theorems about them.  The program of this appendix is an executable realisation of those acts among parties who run it, and the village market of Appendix~\ref{sec:village-demo} runs it: what it establishes is that the four transactions compose into a running village market.  The countersigned transfer of Appendix~\ref{app:digital-more} is a third level, in which the issuer signs every transfer of their own coins: a swap in which $p$ and $q$ exchange coins of $u$ and $v$ is one transaction of Definition~\ref{def:gc-transactions}, and at the third level it needs the signatures of $u$ and of $v$ besides; a redemption that takes an $r$-coin needs $r$'s.  The third level is the secure coins of~\cite{eitan2026securing}: the transactions of Definition~\ref{def:gc-transactions} extended with the issuer's log of every transaction in their coins and with state custodians holding copies of it, the issuer a participant of every such transaction; that paper implements them by communicating volitional agents and proves that the implementation realises the specification at quiescence, conserves money, recovers the log exactly after the loss of a key or a machine, never double-spends after recovery, and finalises a requested payment while the issuer and a supermajority of its custodians stay live.

\mypara{The digital social contract}
The contract of Section~\ref{sec:clauses} is realised by the \emph{digital social contract} implementing the currency: the multiagent program that every participant runs, its instances communicating only with each other~\cite{cardelli2020digital}.  Its enforced clauses are kept by every correct run of the program: the issuer withholds no permitted transaction, their own coins are accepted in payment, redemption proceeds against any coin the issuer holds, of the bearer's choice, and the record makes default public (Section~\ref{sec:insolvency}).  A party also promises, in its undertaking, to run the certified code; that clause is \emph{undertaken}, its record the undertaking in another party's hands, and it is \emph{enforced} where the hosting platform attests the code~\cite{shapiro2026formalising,shapiro2026gsg}.  Mutual attestation for GLP is further work (Appendix~\ref{app:digital-more}), so today the enforced clauses of this section rest on the parties running the certified code.  The pricing undertaking (clause~5) is performed outside the contract.  A person joins an instance of the currency on a friend's invitation, by signing an undertaking to keep the contract towards all its participants and to run its certified code, which the super-app presents to each participant on first contact~\cite{shapiro2026formalising,shapiro2026gsg}.  Enforcement by mutual attestation, default and identity escape, the digital platforms a currency can run on, and privacy are discussed in Appendix~\ref{app:digital-more}.

\mypara{Implementation in GLP}
Grassroots Logic Programs (GLP)~\cite{shapiro2025glp} is a multiagent, concurrent, logic programming language running on Dart for smartphone deployment.  A grassroots specification is realised as a GLP program, which people load and activate voluntarily on their devices.  Here we implement the contract of Section~\ref{sec:clauses} as a program in vGLP~\cite{shapiro2026volition}, GLP with volition guards, with one volition-guarded clause per transaction of Definition~\ref{def:gc-transactions}, and compile it by the vGLP compiler to the GLP program every participant runs, together with the interface constructs the compilation derives from the clauses; the derivation is in Appendix~\ref{sec:derivation}.  The Grassroots Super-App installs the program as a certified mini-app and activates it with its two channels, to its interface and to the social graph; the app renders the compiled program's constructs (Figure~\ref{fig:screens}), and the village market of Appendix~\ref{sec:village-demo} runs on the compiled program with scripted persons in place of the interface.

\subsection{From the Transactions to the Program}\label{sec:derivation}

The program is written in vGLP~\cite{shapiro2026volition}, GLP with volition guards: a clause preceded by a volition guard \verb|*(X1, ..., Xk)| reduces only when the person wills it, and the person's answer fills the writers \verb|X1, ..., Xk|.  The four transactions of Definition~\ref{def:gc-transactions} are four volition-guarded clauses of the program, one each, guarded where the transaction is guarded.  The program is at \repo: the certified mini-app in \texttt{programs/\allowbreak currencies/\allowbreak coins/\allowbreak currency}---its entry \texttt{coins(Id, Ui, Convs)}, the source \texttt{coins\_agent.vglp} and its compilation \texttt{coins\_agent.glp}---and the types and the village market \texttt{village.glp} in \texttt{programs/\allowbreak currencies/\allowbreak coins}.

\mypara{The agent}
Each person runs a \emph{currency agent}, the procedure \verb|agent| of the source, whose state is the person's holdings---the atoms \textcent$(u)$ of Appendix~\ref{app:schemas}, each coin as \verb|coin(Issuer, Serial)|---and the serial its next coin takes.  Its arguments are its name, its own input, the input from its counterparties, the conversations the social graph hands it, its outputs (the screen and one channel per counterparty), its holdings and its serial.  The agent transacts with the agents of other persons over those conversations, and an act arriving from another agent is carried out by the clause that receives it.

\mypara{Mint}
Mint$_p(k)$ is guarded by $p$: a request clause, whose one field is the amount, and whose effect is $c'_p = c_p \cup \text{\textcent}^k_p$:
\begin{verbatim}
*(Amount)
agent(Id, UserIn, PeerIn, Convs, Outs, Holdings, Serial) :-
    ground(Id?), integer(Amount?), Amount? >= 1 |
    create_coins(Id?, Amount?, Serial?, New),
    Serial1 := Serial? + Amount?,
    append_coins(Holdings?, New?, Holdings1),
    show(minted(Amount?), Outs?, Outs1),
    after_change(Id?, UserIn?, PeerIn?, Convs?, Outs1?,
                 Holdings1?, Serial1?).
\end{verbatim}

\mypara{Swap}
Swap$_{p,q}(\text{\textcent}^k_u, \text{\textcent}^{k'}_v)$ is guarded by $p$ and $q$.  At $p$ it is a request clause with five fields---the counterparty, what is given and what is wanted---whose guard is the precondition $\text{\textcent}^k_u \subseteq c_p$, tested by selecting the coins given:
\begin{verbatim}
*(Friend, GiveCoin, GiveAmount, WantCoin, WantAmount)
agent(Id, UserIn, PeerIn, Convs, Outs, Holdings, Serial) :-
    ground(Id?), ground(Friend?), ground(GiveCoin?),
    integer(GiveAmount?), GiveAmount? >= 1, ground(WantCoin?),
    integer(WantAmount?), WantAmount? >= 1 |
    select_coins(GiveCoin?, GiveAmount?, Holdings?,
                 Status, Give, Rest),
    propose_swap(Status?, Id?, Friend?,
                 lot(GiveCoin?, GiveAmount?),
                 lot(WantCoin?, WantAmount?), Give?, Rest?,
                 UserIn?, PeerIn?, Convs?, Outs?, Serial?).
\end{verbatim}
The proposal reaches $q$ with the coins given and a response variable, and at $q$ it is the question of a \emph{responder}, spawned per arriving proposal, whose one volition-guarded clause is the person's acceptance:
\begin{verbatim}
*(yes, From?, Want?, Offered?)
respond_swap(offer(From, Want, Offered), Resp?, Ans?) :-
    ground(From?), ground(Want?), ground(Offered?) |
    accept_answer(From?, Want?, Offered?, Resp, Ans)
 *(no) decline_answer(From?, Offered?, Resp, Ans).
\end{verbatim}
The clause carries an else-branch, which is the decline: an else-branch is enabled by the absence of a will~\cite{shapiro2026volition}, so the refusal fulfils no volition, and it covers the person declining and the proposal left unanswered past the deadline the mediator sets.  The acceptance reaches $q$'s agent as an input, and the clause receiving it tests the precondition $\text{\textcent}^{k'}_v \subseteq c_q$ by selecting the coins wanted: if they are held it binds the response to them and takes the coins offered, which is the effect $c'_q = c_q \setminus \text{\textcent}^{k'}_v \cup \text{\textcent}^k_u$; if not, it declines, returning the coins offered.  The response reaches $p$'s agent as an input, and the clause receiving an acceptance takes the coins given, $c'_p = c_p \setminus \text{\textcent}^k_u \cup \text{\textcent}^{k'}_v$; a refusal returns $p$'s coins to its holdings.  The unbound response variable is the pending consent, and binding it commits or aborts the swap: the transaction is taken only when both guards are willing, as Definition~\ref{def:gc-transactions} states.

\mypara{Pay}
Pay$_{p,q}(k)$ is guarded by $p$ only: a request clause with two fields, the payee and the amount, whose guard is the precondition $\text{\textcent}^k_q \subseteq c_p$:
\begin{verbatim}
*(Friend, Amount)
agent(Id, UserIn, PeerIn, Convs, Outs, Holdings, Serial) :-
    ground(Id?), ground(Friend?),
    integer(Amount?), Amount? >= 1 |
    select_coins(Friend?, Amount?, Holdings?,
                 Status, Give, Rest),
    do_pay(Status?, Id?, Friend?, Amount?, Give?, Rest?,
           UserIn?, PeerIn?, Convs?, Outs?, Serial?).
\end{verbatim}
The coins go to $q$, whose agent takes them without asking its person---clause~3 is the payer's decision---which is the effect $c'_q = c_q \cup \text{\textcent}^k_q$.  Coins that are not $q$'s own are no payment of clause~3, and $q$'s agent returns them.

\mypara{Redeem}
Redeem$_{p,q}(r)$ is guarded by $p$ only: a request clause with two fields, the issuer presented to and the issuer of the coin wanted, whose guard is the precondition $\text{\textcent}_q \in c_p$:
\begin{verbatim}
*(Friend, WantCoin)
agent(Id, UserIn, PeerIn, Convs, Outs, Holdings, Serial) :-
    ground(Id?), ground(Friend?), ground(WantCoin?) |
    take_coin(Friend?, Holdings?, Found, Rest),
    do_redeem(Found?, Id?, Friend?, WantCoin?, Rest?,
              UserIn?, PeerIn?, Convs?, Outs?, Serial?).
\end{verbatim}
The presented coin reaches $q$ with the issuer wanted and a response variable, and $q$'s agent fulfils the presentation without asking its person---clause~4: it takes the coin back and, if it holds an $r$-coin, gives it, which is the effect $c'_q = c_q \setminus \text{\textcent}_r \cup \text{\textcent}_q$; $p$'s agent, receiving the coin, has $c'_p = c_p \setminus \text{\textcent}_q \cup \text{\textcent}_r$.  The other half of the precondition, $\text{\textcent}_r \in c_q$, is $q$'s to test, and where it fails the agent refines the act without departing from it: it returns the presented coin with the issuers of the coins it holds, one per issuer, and the bearer chooses among them; a redemption against the disclosed menu is a choice among the redemptions the menu offers, one of which the issuer then takes.  The disclosed coins are not locked while the bearer chooses, as locking would leave the issuer waiting on a bearer that never answers.

\mypara{Conversations}
Two agents transact over the root channel of the conversation the super-app opens between their two executions of the program, at an invitation of one friend by the other~\cite{shapiro2026gsg}.  The agent takes each conversation from the social graph, adopts its channel, and shows it opened; an act willed towards a party with whom no conversation is open is not sent---the coins stay, and the screen says so.  The program has no network entry and no act beyond the four of the contract: two agents become connected by the social graph's invitation, willed by both friends, and the first act of the contract between them follows---a swap willed by both (Appendix~\ref{app:schemas}, Introduction), or a payment or presentation by a party who already holds the other's coin (Appendix~\ref{app:schemas}, Volition).

\mypara{What the compilation derives}
The vGLP compiler~\cite{shapiro2026volition} produces from the source the GLP program every participant runs, \texttt{coins\_agent.glp}: the agent, in which each volition-guarded clause becomes clauses that pose the clause's question to the person and take its body on the answer, and the \emph{mediator}, which shows each question as an interface construct and routes the answer back.  The constructs are the manifest of the source~\cite{shapiro2026volition}: the four request clauses of \verb|agent| are four compose forms, persistent since the clause is tail-recursive, whose fields are the clause's writers---Mint with its amount; Swap with the counterparty, the coins given and the coins wanted; Pay with the payee and the amount; Redeem with the issuer presented to and the issuer wanted---and the clause of \verb|respond_swap| is one card per proposed swap, showing the proposer, the coins offered and the coins wanted, with the buttons Accept and Decline.  The display declarations of the source name the panel, the label of each form and of each card's Accept, and the widget of each field; the Decline of a card is its else-branch, which fulfils no volition and carries no display declaration of its own~\cite{shapiro2026volition}.  They also declare the agent's report of its holdings as the balances view.  The mediator is emitted into the compiled program, so the program needs nothing beyond itself.

\begin{figure}[t]
\centering
\includegraphics[height=7cm]{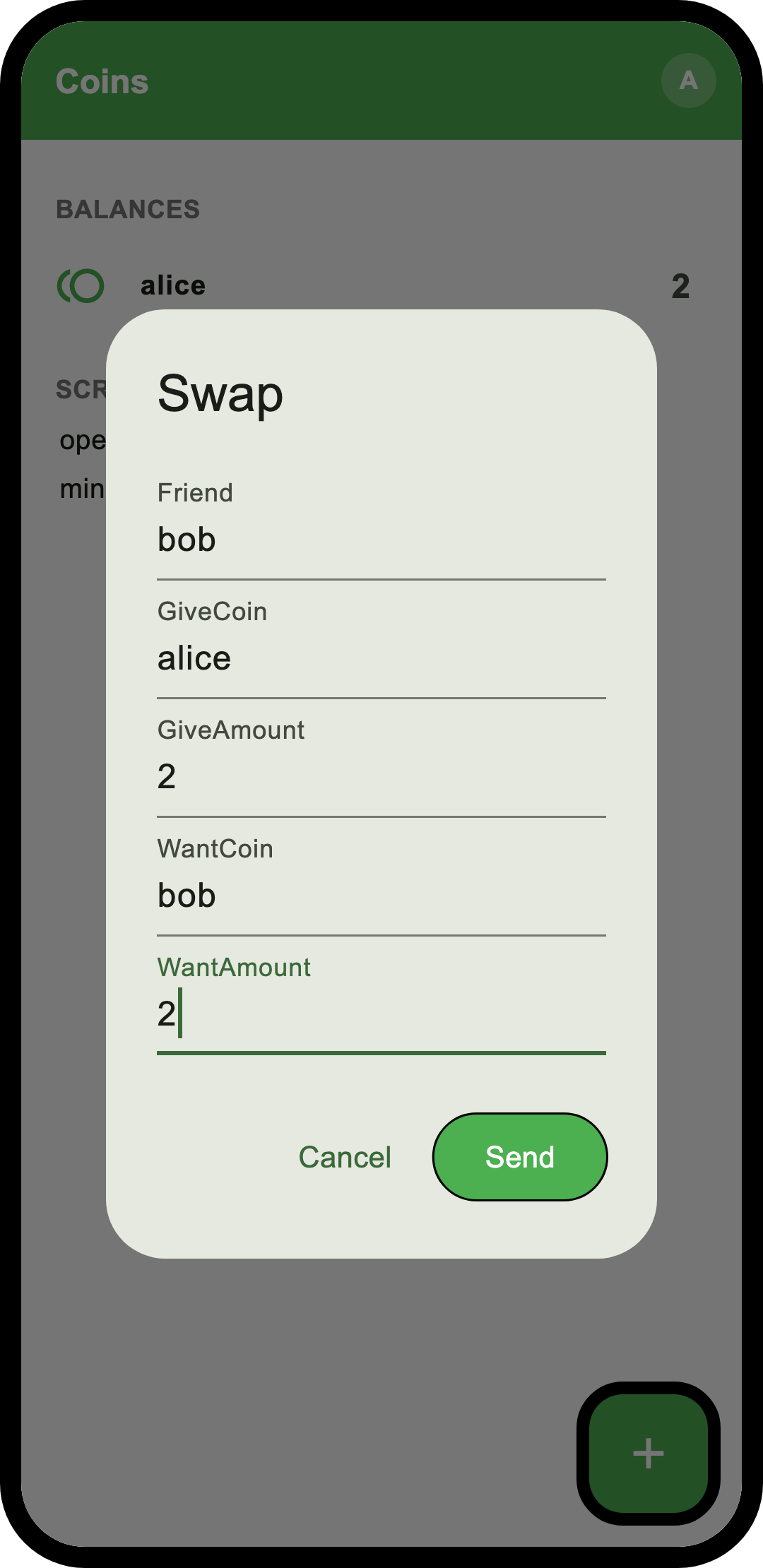}\hspace{6mm}\includegraphics[height=7cm]{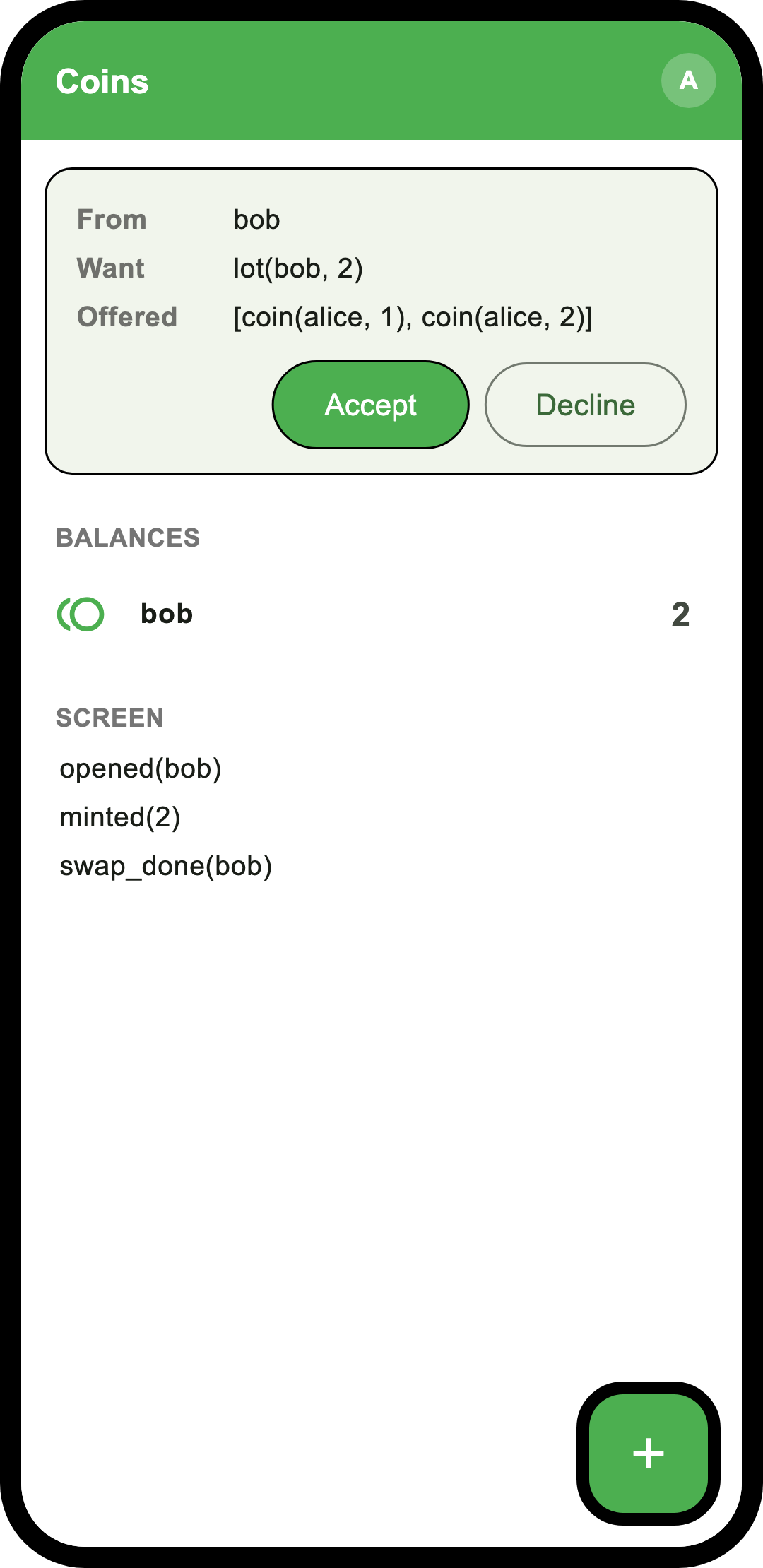}
\caption{The currency mini-app as the Grassroots Super-App renders it from the display declarations of the source: the Swap form at the proposer (left), and the proposal as a card at the counterparty, with Accept and Decline (right).}
\label{fig:screens}
\end{figure}

\subsection{Village Market Demonstration}\label{sec:village-demo}

To demonstrate that the acts of the contract compose into a running village market, we run a village market of six agents on the program: Alice~(baker), Bob~(farmer), Charlie~(carpenter), Diana~(doctor), Eve~(teacher), and Frank~(fisherman).  The scenario exercises six operations, each an act of the contract or a swap of Section~\ref{sec:coins}:

\begin{enumerate}
\item \textbf{Symmetric mutual credit.} Alice and Bob each mint 15~coins and swap them at par, establishing mutual liquidity.  Alice and Charlie do the same with 15~coins each, Charlie and Eve with 10, and Eve and Frank with 10.

\item \textbf{Asymmetric mutual credit.}  Diana gives Bob 20~diana-coins for 24~bob-coins---a premium of four coins for her credit; Diana similarly gives Frank 15~coins for 18~frank-coins.

\item \textbf{Portfolio swap.}  Eve trades 5~frank-coins to Charlie for 5~alice-coins, acquiring the currency she needs to buy bread from Alice.

\item \textbf{Pay.}  Bob pays Alice 5~alice-coins for bread; Eve pays Charlie 6~charlie-coins for a bookshelf; Frank pays Diana 3~diana-coins for a medical checkup; Eve pays Alice 3~alice-coins for bread.

\item \textbf{Redeem.}  Frank redeems 5~diana-coins from Diana, taking back five of his own coins she holds---set-off, reducing his exposure to Diana's currency.

\item \textbf{Sale of debt.}  Alice sells 10~bob-coins to Eve for 4~frank-coins---a discounted sale reflecting Alice's desire to reduce exposure to Bob's debt.
\end{enumerate}

The play \texttt{village.glp} stands in for the super-app and for the persons: it opens the conversation between every two villagers who transact, seven in all, and hands each villager its conversations on its channel to the social graph, and a scripted person on the other end of each villager's channel to the user interface reads the cards the mediator shows and the screen the agent writes, and answers the cards in the order of its script, each act once the one before it has taken effect on its screen.  The run performs the six operations and ends with the holdings of Table~\ref{table:village-holdings}.  They reconcile with Conservation of Money (Lemma~\ref{lem:conservation}): every coin is held against its issuer, and the coins of each issuer held across the six agents are the coins they minted---30 alice-coins, 39 bob-coins, 25 charlie-coins, 35 diana-coins, 20 eve-coins and 28 frank-coins---and are those the six operations produce.

\begin{table}[t]
\centering
\small
\begin{tabular}{@{}ll@{}}
\hline
\textbf{Agent} & \textbf{Holdings after the scenario} \\
\hline
Alice & 5 bob-coins, 15 charlie-coins, 8 alice-coins, 4 frank-coins \\
Bob & 10 alice-coins, 20 diana-coins \\
Charlie & 10 alice-coins, 10 eve-coins, 6 charlie-coins, 5 frank-coins \\
Diana & 24 bob-coins, 13 frank-coins, 8 diana-coins \\
Eve & 4 charlie-coins, 1 frank-coin, 2 alice-coins, 10 bob-coins \\
Frank & 7 diana-coins, 10 eve-coins, 5 frank-coins \\
\hline
\end{tabular}
\caption{The holdings of the six agents at the end of the village market, as each agent's screen reports them, self-held coins included.}
\label{table:village-holdings}
\end{table}
\subsection{Implementation Summary}\label{sec:impl-summary}

The village market of Appendix~\ref{sec:village-demo} is a GLP script that drives the six agents through the six operations.  The implementation is open-source: the mini-app is at \repo, in \texttt{programs/\allowbreak currencies/\allowbreak coins/\allowbreak currency}, and the village market in \texttt{programs/\allowbreak currencies/\allowbreak coins}.

The proof-of-concept represents each coin as a unit; a practical deployment would use range-denominated coins to compress holdings of the same issuer into a single term.  The scenario runs with every party acting as the contract specifies.  Three things are left to further work: an issuer who defaults by refusing a presentation they are obliged to fulfil, and the effect on the liquidity ratios of the holders of their coins; concurrent redemptions contending for the same coins, where the menu above is contested; and populations of hundreds of agents, at the scale of the village of Appendix~\ref{sec:community-bank}.

\section{Deferred Proofs}\label{app:proofs}

This appendix proves Chain Redemption and the solvency, liquidity, and run results of Section~\ref{sec:results}.  The syntactic conditions establishing that grassroots currencies are volitionally grassroots (Corollary~\ref{cor:gc-grassroots}) are checked in Appendix~\ref{app:schemas}.

\smallskip\noindent\textbf{Proposition~\ref{prop:chain-redemption} (Chain Redemption) (restated).} \emph{Let $p_0, p_1, \ldots, p_k$ be a sequence of distinct persons with $k \ge 1$ such that each $p_i$ ($0 \le i < k$) holds a $p_{i+1}$-coin.  Then from that configuration there is a sequence of $k-1$ redemptions by $p_0$ whose effect is that $p_0$ holds a $p_k$-coin.}

\begin{proof}
By induction on $k$.  For $k=1$, $p_0$ already holds a $p_1$-coin: zero redemptions suffice.  For $k>1$, by hypothesis $p_0$ holds a $p_1$-coin and $p_1$ holds a $p_2$-coin.  $p_0$ redeems the $p_1$-coin they hold against the $p_2$-coin held by $p_1$.  After this Redeem $p_0$ holds the $p_2$-coin, and the sequence $p_0, p_2, \ldots, p_k$ has $k-1$ edges with each of its members holding a coin of the next.  By induction, $k-2$ further Redeems put a $p_k$-coin in the holdings of $p_0$.
\end{proof}

\smallskip\noindent\textbf{Proposition~\ref{prop:graph-solvency} (Graph Solvency) (restated).} \emph{The holdings of a sink-closed set $V$ clear into $s$ if and only if $a_p \ge \ell_p$ for every member $p$ of $V$ other than $s$, and cleared holdings leave each such member holding exactly $a_p - \ell_p$ coins of the sink.}

\begin{proof}
A clearing redemption among members of $V$ surrenders a coin of a member $q \ne s$ and takes a coin of $q$'s holdings of an issuer $r \ne q$; if $r \ne s$ then $r \in V$, by sink-closedness of the edge $q \rightarrow r$.  The surrendered coin becomes self-held and the taken coin moves to the redeemer, so every edge the redemption deletes, creates, or moves is incident to $s$ or has both endpoints in $V$; edges outside $V$ are untouched, and $V$ remains sink-closed along any safe computation of clearing redemptions among its members.

Write $n_u := a_u - \ell_u$.  A Redeem in which holder $p$ surrenders a $q$-coin and takes a coin $y$ of issuer $r$ held by $q$ changes the counts as follows.  The surrendered coin moves to its issuer: $a_p$ and $\ell_q$ decrease by 1.  If $r \notin \{p, q\}$, $y$ moves from holder $q$ to holder $p$: $a_q$ decreases and $a_p$ increases by 1.  If $r = p$, $y$ moves to its issuer: $a_q$ and $\ell_p$ decrease by 1.  If $r = q$, $y$ moves from its issuer to $p$: $a_p$ and $\ell_q$ increase by 1.  In every case $n_u$ is unchanged for every person $u$, so every $n_u$ is invariant along any safe computation of Redeem transitions.

Only if: cleared holdings of $V$ have $\ell_p = 0$ for every member $p$ of $V$ other than $s$, so there $n_p = a_p \ge 0$; $n_p$ is invariant, so $a_p \ge \ell_p$ in the given configuration.

If: by induction on $C := \sum_{p \in V \setminus \{s\}} \ell_p$.  If $C = 0$ the holdings of $V$ are cleared.  Otherwise some person $p$ holds a coin of an issuer $q \in V \setminus \{p, s\}$, and $p$ is a member of $V$, as the edge $p \rightarrow q$ is incident to $s$ or lies in $V$.  As $a_q \ge \ell_q \ge 1$, $q$ holds a coin $y$ of some issuer $r \ne q$, and $p$ redeems the $q$-coin against $y$---a clearing redemption among members of $V$.  By the count changes above, the redemption decreases $C$ by 1 or 2 and leaves every $n_u$ unchanged, so the condition persists and the induction applies.  At most $C$ redemptions clear the holdings of $V$.

In cleared holdings of $V$, every coin of another issuer that a member $p \ne s$ holds is a coin of the sink---its issuer is $s$ or in $V$ by sink-closedness, and every issuer in $V$ other than $s$ is cleared---and there are $a_p - \ell_p$ of them, as $n_p$ is invariant and $\ell_p = 0$ there.
\end{proof}

\smallskip\noindent\textbf{Proposition~\ref{prop:clearing-liveness} (Clearing Liveness) (restated).} \emph{If $a_p \ge \ell_p$ for every member $p$ of $V$ other than $s$, then every maximal computation of clearing redemptions among members of $V$ ends with the holdings of $V$ cleared into $s$, after at most $\sum_{p \in V \setminus \{s\}} \ell_p$ redemptions.}

\begin{proof}
A clearing redemption among members of $V$ decreases $C = \sum_{p \in V \setminus \{s\}} \ell_p$ by 1 or 2 and leaves every $n_u$ unchanged (proof of Proposition~\ref{prop:graph-solvency}), so the computation takes at most $C$ redemptions, and $a_p \ge \ell_p$ holds for every member $p \ne s$ in every configuration of it.  Suppose the holdings of $V$ are not cleared in the last configuration: some person $p$ holds a coin of an issuer $q \in V \setminus \{p, s\}$, and $p$ is a member of $V$, as the edge $p \rightarrow q$ is incident to $s$ or lies in $V$.  As $a_q \ge \ell_q \ge 1$, $q$ holds a coin $y$ of some issuer $r \ne q$, and the clearing redemption in which $p$ surrenders the $q$-coin and takes $y$ is enabled there, contradicting maximality.
\end{proof}

\smallskip\noindent\textbf{Proposition~\ref{prop:graph-liquidity} (Graph Liquidity) (restated).} \emph{The liquidity of $A$ is the maximum flow from $A$ to $s$ in $G$.}

\begin{proof}
For the upper bound, for a set $S$ of persons with $A \subseteq S$ and $s \notin S$, let $X_S$ be the number of coins held by a person in $S$ and issued by a person not in $S$; in the given configuration, $X_S$ is the capacity of the cut at $S$.  A Redeem of a member of $A$ never increases $X_S$: if the debtor is not in $S$, the surrendered coin, counted in $X_S$, moves to its issuer, and the coin taken adds at most one; if the debtor is in $S$, the surrendered coin is not counted in $X_S$, and the coin taken, moving from the debtor to the member, is counted after exactly if it was counted before.  Every $s$-coin that a member of $A$ holds in the last configuration is counted in $X_S$, so their total is at most the capacity of every cut, in particular the minimum, which is the maximum flow.

For the lower bound, take a maximum-size set of pairwise edge-disjoint paths from members of $A$ to $s$, each simple.  Along a path $a = u_0 \rightarrow u_1 \rightarrow \cdots \rightarrow u_m = s$ with $a \in A$ and edge coins $e_1, \ldots, e_m$, $a$ redeems $e_1$ against $e_2$, then $e_2$ against $e_3$, and so on: its issuer $u_i$ is not $a$, as the path is simple, and $e_{i+1}$ is held by $u_i$, untouched by the preceding redemptions.  The $m-1$ redemptions move $e_1, \ldots, e_{m-1}$ to their issuers and $e_m$ to $a$, and touch no other coin, so the members execute the paths one after another---a path of length 1 being an $s$-coin its member already holds---and in the last configuration the members of $A$ hold one $s$-coin per path: maximum-flow many.
\end{proof}

\smallskip\noindent\textbf{Proposition~\ref{prop:run} (Run on the Bank) (restated).} \emph{Every run on $p$ is met in full if $a_p \ge \ell_p$, and otherwise ends with exactly $\ell_p - a_p$ $p$-coins in circulation, their presentations unfulfilled.}

\begin{proof}
A redemption of the run surrendering a $p$-coin to $p$ decreases $a_p$ and $\ell_p$ by one each: the surrendered coin becomes self-held, and the coin taken, of an issuer other than $p$, leaves $p$'s holdings.  So $n_p = a_p - \ell_p$ is invariant along the run, and the run takes at most $\min(a_p, \ell_p)$ redemptions.  Such a redemption is enabled exactly when $\ell_p \ge 1$---some person other than $p$ holds a $p$-coin---and $a_p \ge 1$---$p$ holds a coin of another issuer.  If $n_p \ge 0$ then $a_p \ge \ell_p$ throughout, so a redemption is enabled while $\ell_p \ge 1$, and by maximality the last configuration has $\ell_p = 0$.  If $n_p < 0$ then $\ell_p \ge 1$ throughout, so a redemption is enabled while $a_p \ge 1$, and by maximality the last configuration has $a_p = 0$ and hence $\ell_p = -n_p$, the initial $\ell_p - a_p$.
\end{proof}

\section{Liquidity: Worked Example and Recovery}\label{app:liquidity-more}

The material deferred from Sections~\ref{sec:ratio} and~\ref{sec:results}: assets the ratio counts but a lender discounts, a worked example of the ratio and the measures, the special cases of clearing, the run on one member, and recovery from insolvency.

The liquidity ratio counts assets that may not be realisable: foreign coins that cannot be redeemed, coins of insolvent issuers, and fake assets issued by sybil persons to inflate the balance sheet of $p$.  Grassroots currencies are expected to operate over grassroots social networking~\cite{shapiro2023gsn}, where friends reveal their local state to each other, and the local state of a person includes their own neighbourhood: each person sees their friends and their friends' friends.  A lender therefore checks the ratio of $p$ against that neighbourhood: whether the issuers whose coins $p$ holds are persons they can see there, or identities with no standing of their own; and whether the holders of $p$-coins are such persons, or identities of $p$.  Sybil identities cost nothing to create, but a person maintaining mutually-isolated identities in a graph that friends and their friends can see is hard-pressed to do so for long without being exposed~\cite{shahaf2020genuine}; and a mutual credit line is opened only between persons who know each other (Appendix~\ref{sec:implementation}).

\mypara{Worked example}  We compute the liquidity ratio and the net external coin position of the four persons of Figures~\ref{figure:demo1}--\ref{figure:demo4}, staged with paper coins.  In Figure~\ref{figure:demo1}B the green and red persons swapped four coins each, and the green and blue, and the blue and yellow, three each: 20 coins in circulation, every person with a net external coin position of zero and a liquidity ratio of~1.  The trades of Figures~\ref{figure:demo1}C--\ref{figure:demo4}H then move coins as follows: the blue person pays a green coin to the green person (C), a yellow coin to the yellow person (D), and receives two blue coins from the green person (E); redeems their two remaining green coins from the green person against two red coins (F) and pays them to the red person (G); and pays their two remaining yellow coins to the yellow person (H).

At Figure~\ref{figure:demo4}H the holdings are: the yellow person holds three blue coins and has none of their own in circulation; the blue person holds no foreign coin and has four blue coins in circulation, three with the yellow person and one with the green; the green person holds one blue coin and two red coins, and has four green coins in circulation, all with the red person; and the red person holds four green coins and has two red coins in circulation, with the green person.  Hence, from yellow to red, $(a_p, \ell_p)$ is $(3,0)$, $(0,4)$, $(3,4)$, $(4,2)$; the net external coin positions are $-3$, $4$, $1$, $-2$, which sum to zero, as coins are conserved (Lemma~\ref{lem:conservation}); and the liquidity ratios are undefined for the yellow person, who has no coins in circulation and so no liabilities, $0$ for the blue person, $3/4$ for the green person, and $2$ for the red person.  The blue person is insolvent: a run on them leaves all four presentations unfulfilled (Proposition~\ref{prop:run}).  The green person, with $a_p < \ell_p$, is insolvent too, by one coin: a run on them is met for three of the four green coins.  Since all four started the period with a net external coin position of zero, the changes in their net external coin positions over it are those positions themselves.

\mypara{Special cases of clearing}
Proposition~\ref{prop:graph-solvency} has three special cases: (\ia)~$V$ is all persons: the economy clears into the sink exactly when $a_p \ge \ell_p$ for every person but the sink---a liquidity ratio of at least~1 where it is defined, and no condition on a person with no coins in circulation---the sink's own coins staying in circulation; it clears completely exactly when every person holds as much as they owe (Corollary~\ref{cor:complete}), which by conservation of money is the case of equality throughout.  (\ib)~$V$ has no claims on or by the sink: every coin then counts once as an asset and once as a liability within $V$, so the condition holds only with equality throughout---the community clears exactly when every member holds exactly as much as they owe, and each member ends holding only their own coins.  (\ic)~$V$ is a cycle of mutual debt (with the sink an isolated member): the cycle clears with no coins of the sink, as chain redemption around the cycle ends in a set-off.  A sink-closed set is the sink together with any union of connected components of the holding graph without it, so one sink serves many communities, each clearing on its own.

\mypara{The run on one member}
For one member, the liquidity ratio reads: the coins they hold cover the coins they owe, and a run is met in full exactly when the ratio is at least~1 (Proposition~\ref{prop:run}).  Redemption is the bearer's right under the written obligations and transfers no value, so whether a run is met is a property of the holdings themselves, and an economy all of whose members hold at least as much as they owe clears into any member designated as its sink (Proposition~\ref{prop:graph-solvency}, case~\ia).  When $\ell_p > a_p$, the $\ell_p - a_p$ presentations that cannot be met stand unfulfilled, evidencing $p$'s default (Section~\ref{sec:coins}).

\mypara{Recovery from insolvency}
The dynamics of insolvency offers a built-in path to recovery.  The longer a person $p$ is insolvent, the less faith---or patience---the holders of $p$-coins will have for $p$, and hence they may offer to sell $p$-coins as bad debt, at an ever-growing discount.  The lower the price of $p$-coins, the easier it is for $p$ to recover: since $p$ prices their goods and services in $p$-coins, the lower their value, the higher the price $p$ can quote for their goods and services in terms of $p$-coins and still be competitive.  For example, if, following $p$'s insolvency, $p$-coins trade at a 90\%-discount and the market price for mowing a lawn is 11 coins, then $p$ can offer to mow the neighbour's lawn for 100 $p$-coins and still be competitive.  The higher the nominal price of $p$'s goods and services, the faster $p$ can redeem all $p$-coins in circulation, recover from insolvency, and subsequently regain liquidity, at which point the value of $p$-coins returns to market value.  Those who take a hit in such an insolvency-and-recovery cycle of $p$ are the impatient creditors of $p$ who have sold their $p$-coins at a discount; those who trusted that $p$ would recover, and bought $p$-coins at a discount, profit.

\section{Economic Scenarios}\label{sec:community-bank}

The community bank\full{ of \App{sec:bank}}{} in full, and scenarios beyond it: mutual currency recognition, regional banks, Hawala, private banking, corporation currencies, and a community basic income.

\iffull\else\fi

The bank would normally have a governing body and signatories as authorised by the community.  Presumably, community members will be happy to use the community currency, knowing that they own the bank.

\mypara{Community mutual credit scenario}  Consider a village with a community of 501 people in which every two villagers swap 100 personal coins with each other. As a result, each villager will have 50,000 coins of other villagers, while issuing and transferring 50,000 of their own coins to others,  with 25,050,000 grassroots coins in circulation, solely based on mutual trust among villagers and without any external capital or credit.  Such liquidity may jumpstart the village's local economy, with the exposure of each villager to any other initially limited to 0.2\% of the total credit they have issued.

\mypara{Bank--member credit lines}  If a community bank is formed and issues 10,000 community coins to each member in exchange for 10,000 personal coins of that member, each member has 50,000 personal coins of others and 10,000 community coins.  From that point on the community transacts in community coins, unless a member has exhausted their bank credit and must resort to trading with the personal coins they hold, if any.

\mypara{Personal credit lines as backing}  The $p$-coins transferred to the community bank are of equal value to the community coins received by $p$ as long as $p$ is liquid; if $p$ becomes insolvent they lose their value.  The mutual credit lines between community members and $p$ thus back the community coins the bank has provided to $p$, though they are unsecured claims and not collateral in the legal sense: the bank holds $p$-coins, which it may redeem against the coins of other members that $p$ holds.

\mypara{Bank risk management}  Even if the bank initially opens identical credit lines with all community members, it can manage its exposure to members differentially, without withdrawing credit from any of them.  If the bank decides that $p$ poses a higher credit risk than other members, for example than members $q_1, \ldots, q_{50}$, it may redeem from $p$ 5,000 of the 10,000 $p$-coins it holds, in exchange for 100 $q_i$-coins held by $p$, for each $i = 1, \ldots, 50$.  This decreases the exposure of the bank to $p$ by half and increases its exposure to each of the $q_i$ by one per cent, while leaving $p$ with their 10,000 community coins: the bank reallocates part of its risk with $p$ to creditworthy members who trusted $p$ with a mutual credit line, without harming $p$'s liquidity in community coins.  Risk management by redemption is possible only while $p$ is solvent; once it is not, the bank may sell $p$-coins at a discount (Section~\ref{sec:coins}, sale of debt).

\mypara{Mutual currency recognition}  An alternative to forming a community bank is for community members to mutually recognise each other's currencies: $p$ and $q$ agree to accept each other's coins as payment, each up to a per-issuer limit, as agreed by the parties---the same extension, between two parties rather than a community and its bank.  This is a bilateral standing agreement that builds exposure organically as payments flow, without the upfront coin swap of a credit line.  If all community members mutually recognise each other's coins, the community forms a currency area without a community bank.  Similarly, if two banks mutually recognise each other's currencies, their customers can transact across banks directly.  This mechanism enables bottom-up formation of currency areas.

\mypara{Regional banks}  Several community banks from the same region can cooperate.  First, they may establish mutual credit lines, allowing members $p$, $q$ of different communities $a$, $b$ to transact via a two-step chain payment: $p$ redeems from their bank $a$ a $b$-coin, and uses it to pay $q$.  Second, they may form a regional bank $c$ with its own grassroots currency, thus becoming in a sense a single currency area~\cite{mundell1961theory}: if the agreement requires members of the communities $a$ and $b$ to accept $c$-coins, then members of both may transact via $c$-coins, which they can redeem from their respective banks.  Chain redemption ensures an exchange rate at par among personal, community, and regional coins across the region, provided members, communities and the regional bank are mutually liquid.  A condition for admitting a new community bank to the regional bank could be sufficient mutual credit lines between the new bank and the member banks, by which the member banks vouch for the good standing of the new one.  Such grassroots formation of regional banks may continue to span an entire country, resulting in a country-wide grassroots currency even without the support of the central bank.

\mypara{Hawala}
The thousand-year-old trust-based money transfer system of Hawala~\cite{elqorchi2003hawala} can be realised with grassroots currencies.  If $p$ wishes to transfer funds to a remote person $q$, they can do so over the net; but $p$-coins may be of no use to $q$ if there is no liquidity between $p$ and the merchants in $q$'s locale.  A Hawala-like network uses the extension of \App{sec:bank} instead, in voluntary chain transfers: $p$ transfers locally-liquid coins to their local Hawala agent, and $q$ receives from their local Hawala agent the same number of coins with liquidity in $q$'s locale; credit is settled among the Hawala agents along the path of transfers, as it has been for a thousand years.

\mypara{Private banking}
A member $b$ of the community with initial capital may aim to operate as the community banker, following the footsteps of the Medicis and the Rothschilds.  By opening mutual lines of credit with all creditworthy members of the community, the banker's grassroots currency can become the de facto community currency, as all transactions in the community can be completed by a two-step chain payment via the banker: if $p$ wishes to pay $q$---both with mutual liquidity with the banker $b$---then $p$ can redeem a $b$-coin they hold in exchange for a $q$-coin the banker holds, and then use it to pay $q$.  Naturally, the banker may charge transaction fees.  Still, a resourceful community with mutual trust among its members may sidestep the Medicis and the Rothschilds and establish a community bank, owned, operated and governed by the community.

\mypara{Corporation currencies}
In the past, companies with captive employees (e.g., remote mining or logging towns) paid employees in company currency termed scrips, with which the employees could purchase goods at company-owned shops, resulting in the company controlling the entire economy of the town~\cite{green2011company}.  With grassroots currencies, the situation is the opposite: if a corporation and its employees both use grassroots currencies, then the corporation should pay an employee's salary using the employee's coins.  More generally, a corporation, or a cooperative, may issue its own grassroots currency and price its goods and services in terms of that currency.  People who wish to purchase goods and services from the corporation would then have to obtain company currency, via a mutual credit line, by redeeming it from others, or by purchasing it from the corporation at market price.  Once customers hold company currency they are captive, and a smart company would do its utmost to keep them and have them purchase as much company currency as possible.  A corporation should strive to establish mutual credit lines with its employees, suppliers and customers, providing it with liquidity without external capital or bank credit; such mutual credit lines, if within a given community, could also function as collateral for a mutual credit line between the corporation and the community bank.

\mypara{Community basic income}
A community bank may offer a community-based basic income~\cite{assiagood,circles-UBI,trustlines,howitt2019roadmap}.  To do so, the bank can increase the mutual credit line with members over time, say by one coin a day for each member, even if the member is not creditworthy.  The bank thus provides a form of basic income for those who need it, and a kind of basic pension fund for those who do not need additional credit at present.  To offset write-offs for members who become insolvent or perish while having a negative net investment position, and to maintain its capital adequacy, the community bank will need contributions from inside or outside the community, in the form of philanthropic donations, government support, or increased income from fees.

\section{Digital Realisation, Continued}\label{app:digital-more}

The discussion deferred from Appendix~\ref{sec:implementation}: the issuer as the sole credit risk, the digital implementations, the legal status of digital coins, enforcement by mutual attestation, default and identity escape, digital platforms, and privacy.

\mypara{The issuer is the sole credit risk}
A grassroots coin is its issuer's debt, so bearing the issuer's credit risk is unavoidable.  A paper coin, transferred by the holder signing it over and delivering it (Section~\ref{sec:paper-coins}), makes every past holder one too~\cite{boe1882,ucc3}.  The design reduces credit risk to the issuer; the correctness of the code, the custody of keys, the availability of the issuer and of the custodians, and the certification of the compiler are trust assumptions of their own.  For an implementation to add no counterparty beyond the issuer, we require that the issuer $p$ sign every transaction in a $p$-coin, and that $p$ may not withhold their signature from a transaction that the protocol permits: a payment or redemption of a $p$-coin by its holder, or a swap the holders of the coins have agreed.  Then holdership of a $p$-coin is proven by $p$'s signatures, and the only counterparty trusted is $p$, not to disown their signature---the trust already given when accepting $p$'s debt.  If $p$ signs conflicting transfers of the same coin (double spending, or equivocation), $p$'s own signatures prove the fraud.  On paper, the requirement is met by the issuer countersigning each transfer, naming the new holder: holdership is then proven by the issuer's signature, and nothing more is needed than paper, a pen, and the issuer's signature on each transaction. Note that while $p$ may double-spend a fresh $p$-coin, this irrational act amounts to economic suicide: it is a signed statement by $p$ that $p$ is not creditworthy, and it results in no financial gain for $p$, as $p$ could have issued more $p$-coins instead of double spending.  A recipient doubting a received $p$-coin may redeem it from $p$ against a fresh $p$-coin, upon which $p$ either provides such a coin or provides evidence of the double spend.

\mypara{Digital implementations}
The issuer countersigning every transfer of their coins is the design of Grassroots Flash~\cite{lewis2023grassroots}, a payment-system protocol for grassroots currencies: a payment in a currency is final when the issuer of the currency approves it, so finality is definite and no consensus is needed.  \full{A separate paper}{Another paper}~\cite{eitan2026securing} makes the loss of a key or machine recoverable: identity and state custodians designated by the person restore both, with the record of their currency recovered exactly and its issuer resuming without double spending.  Disowning a key is then a deliberate act, not an accident.  We implement this for a currency: the issuer keeps an append-only log of the mints and payments in their own coins, numbered in order; a state custodian reads the log and acknowledges each entry, and the parties take a transaction as final once it is acknowledged.  A copy of the log is kept, and the issuer recovers from it the number they resume from.  The code is at \repo, in \texttt{programs/\allowbreak currencies/\allowbreak bonds\_v2/\allowbreak secure}, a single-custodian version of the log; the custodian set, finality by a supermajority of custodians, and the recovery of a person's identity and holdings are specified, implemented and proved in~\cite{eitan2026securing}.  The decisions behind the code are in Appendix~\ref{app:impl-notes}.

\mypara{Digital coins}
A grassroots coin can be implemented as a cryptographically-signed digital IOU: the issuer's identity is their signature key.  In the United States a signature, contract or record may not be denied legal effect, validity or enforceability solely because it is in electronic form~\cite{esign2000}; in the European Union an electronic signature shall not be denied legal effect and admissibility as evidence solely on the grounds that it is in electronic form~\cite{eidas2014}; and the UNCITRAL-model acts say the same where they are adopted~\cite{uncitral2001esignatures}.  Electronic form therefore presents no obstacle of the kinds these statutes address, and the ordinary rules of contract formation and enforceability remain applicable: a digital grassroots coin is a signed, transferable IOU of its issuer.  Three propositions are distinct and only the first follows from the statutes above: that the record is admissible and not denied effect for being electronic; that the obligation it states is an enforceable debt, which depends on the ordinary rules and on the jurisdiction; and that the instrument is a negotiable note within the bills and notes acts~\cite{boe1882,ucc3,geneva1930}, which it does not claim, a grassroots coin promising no sum certain in money.  The protocol records the provenance of each coin and prevents its double spending~\cite{lewis2023grassroots}, serving the purpose that transfer and proof-of-holdership rules serve for paper notes; the record serves that purpose whatever status a jurisdiction grants it.  Identities are self-chosen keypairs, so fictitious issuers---Sybil identities~\cite{douceur2002sybil}---cost nothing to create.  Such an identity issues debt with no credit standing of its own; which legal person the debt is attributable to, and whether issuing it was fraudulent, depend on the facts and on the applicable law.  Grassroots currencies are thus cryptocurrencies, resting on the same cryptographic foundations as the global ones; the substantive distinction is between \emph{global} cryptocurrencies---Bitcoin, Ethereum---and \emph{grassroots} cryptocurrencies (Appendix~\ref{sec:related-work}).

\mypara{Enforcement by mutual attestation}
A digital social contract is a voluntary agreement---but once adopted, participants must not be able to deviate from the contract as programmed.  Smart contracts achieve this through global consensus among third-party validators on a blockchain; the grassroots currencies transactions involve only bilateral interactions, with no need for global agreement on the order of events, so safety reduces to ensuring that each participant is running the correct code, while liveness needs the agents a transaction names to be available besides.  This can be achieved by mutual attestation: when two agents communicate, each proves to the other that the message was produced by a correct execution of the agreed-upon contract module.  The threat model intended is that an attested agent either executes correctly or crashes, rather than deviating arbitrarily~\cite{lamport1982byzantine}.  The contract runs on the participants' own devices, and compliance is verified bilaterally.  Realising mutual attestation for GLP, and establishing that model of it, are left to further work.  Until it is realised, a party who runs other code is in breach of their undertaking (Appendix~\ref{sec:implementation}) and is not prevented by the platform.

\mypara{Default, disappearance, and identity escape}
As in real life, indebted people can default, disappear, or change identity.  An insolvent person may opt to escape their obligations by abandoning their digital identity and starting afresh---not unlike bankruptcy, but unregulated.  Bankruptcy is a regulated process that aims to mitigate the damage to creditors, whereas shedding one's digital identity and assuming another is akin to fleeing the country in order to escape creditors.  To avert that, a mechanism to deal with duplicate identities (Sybils) of the same person is needed~\cite{shahaf2020genuine}: if the holder of $p$-coins discovers that $p$ and $p'$ are identities of the same person, then they may rightfully try to redeem their $p$-coins from $p'$.  Any individual faces death, after which they cannot consume or produce; the net investment position of a person upon death determines whether the holders of their coins can redeem them or suffer a loss.  We leave open the question of inheritance and the management of the estate of a grassroots currency account, noting that accounts held by several signatories are suited to any method devised to address it.

Community-level risk management (\App{sec:bank}), and the collateral and escrow of grassroots bonds~\cite{shapiro2026bonds}, provide measures to mitigate such risks, just as their traditional counterparts do.  In a grassroots digital economy, mutual credit lines serve a dual role: they provide liquidity, and they back the credit community banks extend.  The credit granted to a community member by their peers offers the best mechanism of sybil-resilience, since mutual credit lines are established between people who know and trust each other, making the risk of a correct person opening a mutual credit line with a fake identity low.

\mypara{Digital platforms}
Digital IOUs can be realised on any of the existing digital platforms~\cite{shapiro2025characterising}: centralised, with an operator such as a bank or the sovereign storing the record; decentralised, with smart contracts on a blockchain storing it; federated, with community and institution servers storing it for their members; and grassroots, with no global infrastructure and no third party between any two agents---each person storing their own record~\cite{lewis2023grassroots}.  

On a grassroots platform, a payment involves no one but the payer, the recipient, and the issuer of the coins paid---the recipient and the issuer being one party under clause~3, and two under the extension of \App{sec:bank}: the three can transact whenever they can communicate---offline included---instantly and at no cost, regardless of the state of everyone else. Offline covers physical proximity and local networks: nearby phones can communicate directly over Bluetooth, and phones on the same local network can communicate over it even when the network is disconnected from the wider Internet. The other classes place a third party inside every payment---the operator, the consensus nodes and their fees, the parties' servers---so each payment bears that party's availability and cost.

\mypara{Privacy}
A payment employs the public keys of its parties, so it can be encrypted and known only to them.  The provenance of a coin must be visible for the payee to check its integrity, and all transactions in $p$-coins eventually become visible to $p$, who countersigns them.  Still, privacy from $p$ can be maintained by anonymous sub-accounts: a person $q$ who issues a currency must have their public key known to their social and economic circle, but may maintain an additional anonymous identity $q'$ that does not issue a currency, used by $q$ to send and receive payments indirectly.  Privacy veils may have to be lowered among persons who offer each other mutual credit lines, as each should be aware of the financial state of the other---in line with common business practice, and with grassroots social networking, upon which grassroots currencies are expected to operate: friends reveal their local state to each other~\cite{shapiro2023gsn,lewis2023grassroots}.

\section{Implementation Notes}\label{app:impl-notes}

The decisions behind the secure-currency code of Appendix~\ref{sec:implementation}, at \repo, in \texttt{programs/\allowbreak currencies/\allowbreak bonds\_v2/\allowbreak secure}.  Two things they depend on.  GLP admits one writer and one reader occurrence of a variable, and a reader occurrence in a guard does not count~\cite{shapiro2025glp}.  And each entry of the issuer's log carries, besides its transaction, references to the acknowledgements the issuer has seen---its \emph{tips}---so the log and the acknowledgements are interlaced.

\begin{enumerate}
\item \emph{The first entry carries no tips.}  The issuer's tips are the custodian's acknowledgements and the custodian's acknowledgements are of the issuer's entries, so the first entry is produced by a procedure of its own, with no tips, and the loop starts from the second.

\item \emph{The acknowledgement stream is split rather than read twice.}  It has two readers---the collection of tips, which feeds them back to the issuer, and the binding of finality---so it is split in two, under the relaxation that admits two reader occurrences of a ground value.

\item \emph{A payment carries one reply variable per party.}  The payer and the payee each get an unbound variable the issuer binds, rather than sharing one, so each has a single writer and a single reader.

\item \emph{Finality is bound from a parallel stream read in step with the acknowledgements.}  The reply variables are taken out of the approval requests into a stream of pending entries, and bound when the acknowledgement of the matching number arrives; both streams are in entry order, so they stay aligned.

\item \emph{Tips are collected without waiting.}  The acknowledgement stream is read on the test for an unbound tail: an acknowledgement that has arrived is taken as a tip, one already taken is skipped, and an empty stream yields none.  Producing an entry therefore never waits on the custodian.

\item \emph{Recovery is a second entry point, not a change to the first.}  The wiring that keeps a copy of the log is separate, so the run without recovery is untouched by it.

\item \emph{The copy of the log is taken from the issuer's own stream.}  The custodian consumes entries and emits acknowledgements, and returns no copy; the copy stands for what a custodian would hold.

\item \emph{The number is recovered from a closed stream.}  The copy closes when the issuer runs out of requests, and walking it then yields the number of the last entry, or zero for an empty log.

\item \emph{Recovery relies on the log only growing.}  Any copy of it is a prefix, so the number it ends at is one the issuer may resume from.

\item \emph{The recovered number is compared in a guard.}  A reader occurrence in a guard does not count, so the comparison leaves the single-reader rule intact.

\item \emph{Tips are collected from a list of acknowledgement streams.}  There is one custodian, and a second is a longer list rather than a change to the collection.
\end{enumerate}

\section{Related Work, Continued}\label{sec:related-work}

The comparisons of Section~\ref{sec:related}, in full.

\subsection{Global Cryptocurrencies}\label{sec:crypto}

Global cryptocurrencies such as Bitcoin~\cite{nakamoto2008bitcoin} and Ethereum~\cite{buterin2014next} are unbacked digital assets whose scarcity and transaction cost are controlled by the protocol.  Grassroots currencies differ from global cryptocurrencies in their foundations, architecture, and the financial instruments they support.  This section develops these contrasts along four axes: architectural design, source of value and liquidity, the nature of the financial instruments each system supports, and the attacks each design is open to.

\mypara{Architecture}
Global cryptocurrencies maintain a single replicated ledger---a totally-ordered blockchain---via consensus among miners or validators.  This architecture entails well-known costs: proof-of-work entails high-cost computing~\cite{nakamoto2008bitcoin}; proof-of-stake requires capital lockup~\cite{wood2014ethereum}; and a validating node maintains a copy of the global state.  Transaction throughput is bounded by block size and block interval, and transaction fees arise from competition for inclusion in the next block.

Grassroots currencies employ a different architecture.  Transaction cost is limited to operating a networked device.  The system is grassroots~\cite{shapiro2024grassroots}: multiple independent deployments can emerge without global resources and interoperate once interconnected.

The comparison table in~\cite{lewis2023grassroots} details 18 dimensions along which the two architectures differ, including: one cryptocurrency per blockchain vs.\ one per person, mutually pegged; competitive block creation vs.\ decentralised and cooperative; global all-to-all dissemination vs.\ local grassroots dissemination; probabilistic finality vs.\ definite finality; and a business model based on ICOs and miner remuneration vs.\ a public good.

\mypara{Source of value and liquidity}
Bitcoin and Ether derive their value from market demand; they are unbacked, and liquidity is obtained by mining or purchasing with external capital.  Fiat-backed stablecoins are a class apart, backed by reserves.  Grassroots currencies are closer to `inside money'~\cite{cavalcanti1999inside}---a medium of exchange backed by private credit.  Each person backs their coins with their offerings: goods and services for people and corporations; interest and transaction fees for banks; taxes for municipalities; and fiat currencies for central banks.  Liquidity arises not from capital injection but from mutual credit lines among persons who trust each other, and coin redemption---the obligation to redeem any coin one has issued against any coin one holds---pegs mutually-liquid currencies at par.  Grassroots bonds~\cite{shapiro2026bonds} extend this foundation with interest-bearing credit.

\mypara{Financial instruments on blockchains}
The programmability of blockchains via smart contracts~\cite{szabo1997formalizing,buterin2014next,scilla2018} has given rise to Decentralised Finance (DeFi)~\cite{werner2022sok,schar2021defi}, which re-implements traditional financial instruments on-chain.  Lending protocols~\cite{gudgeon2020loanable,leshner2020compound} offer overcollateralised loans with algorithmically-determined interest rates; decentralised exchanges provide token trading via automated market makers; and stablecoins offer price-stable media of exchange.

Mutual credit and the sale of debt are voluntary swaps of grassroots coins (Section~\ref{sec:coins}), without a blockchain or consensus, and the instruments built on grassroots bonds---loans, credit lines, options---are voluntary swaps of bonds and escrow agreements~\cite{shapiro2026bonds}.  Two differences are noteworthy.  First, many permissionless DeFi lending protocols use overcollateralisation enforced by smart contracts, counterparty identity being pseudonymous and trust absent; grassroots credit is extended among persons with mutual trust, and collateral is available but not mandatory.  Second, DeFi instruments incur gas fees for every state transition; grassroots transactions incur no fees beyond bilateral communication.

\mypara{Crypto-native constructs}
Three constructs of the blockchain architecture rest on properties grassroots currencies do not have---a globally shared, ordered and atomic state---so the attack surfaces characteristic of them do not arise from the base protocol here: flash loans, automated market makers (AMMs), and maximal extractable value (MEV).  A comprehensive systematisation analyses 181 real-world DeFi attack incidents~\cite{zhou2023sokattacks}.

\emph{Flash loans}~\cite{qin2021attacking} are uncollateralised loans that exist within a single atomic blockchain transaction: a borrower takes a loan, uses the funds across arbitrarily many smart contracts, and repays within the same transaction; if repayment fails, the entire transaction reverts.  This is possible because blockchain transactions are atomic across the global state.  Grassroots currencies have no global atomic transaction spanning many parties, so a loan taken and repaid within one such transaction has no counterpart in the base protocol.  Flash loans are a major attack vector against DeFi protocols, with documented exploits yielding returns exceeding 500,000\%~\cite{qin2021attacking}; the base protocol offers no such transaction to exploit.

\emph{AMMs} provide algorithmic price discovery via pooled liquidity and a bonding curve (e.g., the constant product formula $x \cdot y = k$).  On a blockchain the liquidity pool is globally-shared state maintained by consensus.  Grassroots currencies have no pooled global state in the base protocol; price discovery is by bilateral negotiation.  The trade-off is that grassroots currencies lack automated price discovery but have no pool, so neither impermanent loss, the exposure of a liquidity provider, nor the concentration of value in pool contracts that creates targets for exploitation, and no transaction order for sandwich attacks~\cite{daian2020flash}.

\emph{MEV}~\cite{daian2020flash} is the value extracted through control of the inclusion, exclusion and ordering of transactions in a block.  MEV manifests as frontrunning, backrunning, and sandwich attacks, and has been estimated to extract hundreds of millions of dollars annually from Ethereum users~\cite{qin2022quantifying}.  MEV is an artefact of a block producer ordering a global sequence of transactions.  Grassroots currencies have no global ordering and no block producer---each person's record is their own---so extraction by ordering does not arise in the base protocol.

\mypara{Summary}
Grassroots currencies occupy a different point in the design space than global cryptocurrencies and their DeFi instruments.  Where global cryptocurrencies operate among strangers through consensus at the cost of high transaction fees, capital requirements, and exposure to flash-loan attacks, MEV extraction, and sandwich attacks, grassroots currencies operate without consensus through bilateral trust at the cost of requiring social relationships for liquidity.  The financial instruments built on grassroots bonds are comparable in expressiveness~\cite{shapiro2026bonds}, but differ in their trust assumptions, cost structure, and the attacks they are open to.

\subsection{Finance}\label{sec:finance}
Here we survey additional related work: trust-based lending, informal community finance, financial contract specification, credit networks, IOU-based currencies, mutual credit systems, promissory notes, off-chain payment protocols, and monetary theory.

\mypara{Microfinance and peer-to-peer lending}
The Grameen Bank model demonstrated that trust-based, collateral-free lending can work at scale via group liability~\cite{morduch1999microfinance}.  Online peer-to-peer lending platforms extended this principle digitally: Lin et al.~\cite{lin2013friendship} showed that friendship networks on Prosper.com function as credible credit signals, increasing funding probability and predicting lower default rates; Iyer et al.~\cite{iyer2016screening} found that peer lenders predict defaults 45\% more accurately than credit scores.  Grassroots currencies formalise this trust infrastructure: where microfinance relies on group liability and P2P platforms on centralised matching, grassroots currencies encode bilateral trust directly as mutual credit lines, with creditworthiness assessable via the measures of Section~\ref{sec:insolvency}.  This is relevant to financial inclusion: the World Bank Global Findex~\cite{demirguckunt2022findex} estimates that 1.4 billion adults worldwide lack access to formal financial services, often in communities where mutual trust is abundant but capital is scarce---the conditions under which grassroots currencies can bootstrap a local digital economy.

\mypara{Informal community finance}
Rotating savings and credit associations (ROSCAs)---known as chit funds, tontines, or \emph{susus} across cultures---are the dominant informal financial instrument in developing economies.  Besley et al.~\cite{besley1993roscas} proved formally that both random and bidding ROSCAs yield higher utility than autarky by enabling earlier access to indivisible goods through community-based mutual commitment, without formal institutions.  Ardener~\cite{ardener1964roscas} provided the foundational cross-cultural survey documenting their global prevalence.  Hawala networks~\cite{elqorchi2003hawala} transfer value across borders on pure trust, without formal banking infrastructure.  All three---ROSCAs, hawala, and grassroots currencies---share the same foundation: bilateral trust within a community substitutes for institutional intermediation.  Grassroots currencies can be viewed as a digital formalisation of the same principle, and grassroots bonds~\cite{shapiro2026bonds} extend it with the expressiveness of a bond market.

\mypara{Financial contract specification}
The foundational work on composable financial contract specification predates blockchain: Peyton Jones et al.~\cite{peytonjones2000composing} introduced a combinator library for describing financial contracts declaratively, inspiring subsequent blockchain domain-specific languages.  Marlowe~\cite{seijas2018marlowe}, Cardano's financial contract DSL, specifies instruments declaratively with exhaustive static analysis, for execution on a blockchain; Scilla~\cite{scilla2018} is an intermediate-level smart-contract language, automata-based, for verified contracts on a blockchain.  Both run under blockchain consensus.  The Accord Project~\cite{accord2026faq} writes a legally binding agreement as natural-language text with machine-executable components---a template, a data model and logic---and is blockchain-agnostic, a blockchain not being necessary; its clauses execute against external systems, where a grassroots social contract is compiled to a protocol its parties run.  Grassroots currencies take the same declarative approach---the currency is specified by a grassroots social contract (Section~\ref{sec:clauses}), and every transaction is an act of it---but enforce it by the participants' execution rather than via consensus, with each issuer guaranteeing the security of transactions in their currency~\cite{lewis2023grassroots}.  To the best of our knowledge, no prior work specifies a currency by a social contract in natural language, derives from it the transactions of a multiagent protocol, and realises them by a program its participants run with no global execution layer.

\mypara{Credit networks}
Credit networks~\cite{dandekar2011liquidity} study liquidity in networks of bilateral credit lines, with chain payments as the key instrument.  Grassroots currencies inherit this structure: mutual credit lines formed via coin swaps, together with coin redemption, provide for chain payments as in credit networks.  Much of the body of knowledge regarding liquidity in credit networks carries over to grassroots currencies, with the caveat that lack of liquidity not only causes chain payment failures but also devalues the currencies of the liquidity-constrained parties involved.

\mypara{Clearing in financial networks}
Eisenberg and Noe~\cite{eisenberg2001systemic} model an economy of interbank obligations and prove that a clearing payment vector---every node paying its creditors pro rata out of what it holds and receives---exists as a fixed point; the model grounds the analysis of contagion and systemic risk in financial networks~\cite{glasserman2016contagion}. Grassroots clearing differs: obligations are unit coins, redeemed in kind at the holder's initiative rather than paid pro rata in a common num\'eraire; whether a community clears is a comparison of counts, person by person (Proposition~\ref{prop:graph-solvency}), with the sink's coins the residual holding; and what a group of holders can realise now is bounded by a maximum flow (Proposition~\ref{prop:graph-liquidity}).

\mypara{Liquidity as network flow}
Ghosh, Mahdian, Reeves, Pennock and Fugger define a trust network as a decentralised payment infrastructure in which payments are routed as IOUs between trusted entities, over directed links whose capacities are the credit one agent extends another; what a set of agents can pay one of them is the maximum flow from the set to it, a budget constraint on the set~\cite{ghosh2007trust}.  Karlan, Mobius, Rosenblat, and Szeidl define network-based trust as the largest amount one agent can borrow from another, a maximum flow of social collateral over trust links~\cite{karlan2009trust}; in credit networks, a payment goes through exactly when the maximum credit flow from payer to payee covers it, the model behind Ripple-style path-based settlement~\cite{dandekar2011liquidity}.  In all three, the edges are capacities---permissions to borrow or to route---and the question is point-to-point feasibility.  Here the edges are coins already outstanding, the sink is the designated member, and the maximum flow is the number of their coins a group can obtain by redemptions (Proposition~\ref{prop:graph-liquidity}).

\mypara{IOU credit networks and personal currencies}
Fugger's proposal that money can be represented as IOUs in social trust networks~\cite{fugger2004money} is an early articulation of the idea underlying grassroots currencies.  The later XRP Ledger~\cite{schwartz2014ripple} put IOU routing on a ledger with global consensus.  SilentWhispers~\cite{malavolta2017silentwhispers} proposed a distributed, privacy-preserving credit network that requires no ledger, but it addresses only payment routing, not the richer financial instruments that grassroots bonds support.  Holochain~\cite{harrisbraun2024holochain} is an agent-centric framework with no global consensus: each agent keeps its own hash chain, and its white paper describes a currency as accounting records on the chains of the parties, a transfer is one entry countersigned by spender and receiver, and a double spend is a fork of the spender's chain, exposed by the application's shared database or remedied socially.  A grassroots coin is its issuer's debt in the issuer's own unit, transferable and redeemable as of right, and its issuer countersigns every transfer of it.  Hypersyn~\cite{ramabaja2022hypersyn} is a peer-to-peer mutual credit protocol with no consensus algorithm, no distributed ledger and no validators: each node creates its own credit, an edge between two nodes holds reserves of both credits, a payment adds the payer's credit to the edge and removes the payee's by a constant-product rule, so the exchange value of a node's credit is set by the reserves on its edges and updated by arbitrage around triangles of nodes; a node cannot hold another node's credit, so nothing is transferred and nothing is redeemed.  A grassroots coin is held, transferred and redeemed, and priced at par by its issuer's undertaking. Grassroots currencies implement Fugger's vision without blockchain: personal coins are IOUs traded via bilateral transactions, with coin redemption pegging mutually-liquid currencies at par.  Grassroots bonds extend this by adding maturity dates, enabling interest-bearing credit.

\mypara{Mutual credit and community currencies}
Mutual credit systems are the closest precedent to grassroots currencies' approach.  LETS, the most widely implemented, are pure accounting systems of exchange: each member has an account with a central administration unit recording transfers in units, a member creates money by going into debt, and all accounts sum to zero; their stability rests on credit limits, monitoring and sanctions against over-issue~\cite{schraven2001mutual}.  Grassroots currencies have no common unit and no administration: a coin is the debt of its issuer to its holder, in the issuer's own unit, and the issuer alone backs it.  The Swiss WIR, founded in 1934, is the largest and oldest exchange based on a private form of money: a bank keeping an account for each member firm or household in credits denominated in, but not redeemable for, Swiss francs, with a limit on the negative balance; its turnover rises when ordinary money is short, so it is counter-cyclical~\cite{stodder2009wir}.  A systematic review of community currencies finds that they contribute mostly to social sustainability, their economic benefits limited by their small scale~\cite{michel2015community}.  Sardex, a mutual credit network in Sardinia, achieves significant real-economy impact; its cyclic transaction structure has been analysed in \emph{Nature Human Behaviour}~\cite{iosifidis2018sardex}.  The Sarafu community currency in Kenya~\cite{mattsson2022sarafu}, one currency administered by a foundation with its transactions duplicated onto a public blockchain, is an academically-studied digital community currency deployment.  Sardex creates liquidity from trust within a community without external capital---the same principle underlying grassroots currencies---but relies on a centralised operator, and a member's debt or credit is not bilateral towards another member but towards the circuit as a whole, on a credit line the circuit manager sets~\cite{fleischman2020liquidity}; Sarafu relies on its administering foundation.  A grassroots coin is the debt of one person to another.

Circles~\cite{circles-UBI,circles2026docs} implements blockchain-based personal currencies: every account mints its own at a fixed rate, with demurrage, and the currencies are swapped one-for-one along the social trust graph, so a payment reaches a stranger through a path of trust.  Its trust relation authorises the exchange; a grassroots holding is a debt already extended, on which redemption acts.  The Trustlines Network~\cite{trustlines} implements mutual-credit networks on blockchain infrastructure, with bilateral credit lines similar to those of grassroots currencies but under blockchain consensus.

\mypara{Personal currencies}
FairTrade~\cite{hu2012fairtrade} is an incentive scheme for peer-to-peer file sharing in which each peer issues a personal currency, spent by its issuer or by any peer who received it; the issuer must accept its own currency as payment for its services, which is what gives it value, and every other peer accepts it up to a credit limit it sets the issuer by the issuer's trading history, with no central bank or authority.  Its currencies are claims on the issuer's services, and a holder has no right to redeem one against what the issuer holds.  The idea of a personal cryptocurrency was floated around a decade ago~\cite{delarouviere2013personal}.  Promify~\cite{promify2022} enabled artists and celebrities to issue their personal coins and sell their fans exclusive or preferred access to content or merchandise in exchange for these coins---a proof of the viability of personal currencies, yet controlled by a third party.  Worldcoin~\cite{worldcoin2022}, which aims to offer a single global currency for the multitudes, lies at the antipode of grassroots currencies: it is one global currency, where grassroots currencies are one per person.  Prior work includes mutual credit, personal currencies, payments along paths of trust, routing without a ledger, agent-centric execution, and executable legal contracts.  Grassroots currencies are transferable person-issued debts redeemable as of right, specified by a social contract compiled to a program every participant runs, with no shared ledger and no global consensus, and with exact conditions on the holdings for clearing, liquidity and a run on the bank.

\mypara{Bills of exchange and promissory notes}
Before banks dominated payment, debts circulated as bills of exchange and promissory notes---signed promises to pay---traded and discounted~\cite{rogers1995early,santarosa2015financing}.  A grassroots coin is analogous to such an instrument, in digital form: a signed, transferable debt payable on demand, sold at a discount (sale of debt, Section~\ref{sec:coins}), with the protocol serving the purpose that transfer and proof-of-holdership rules serve for paper notes (Appendix~\ref{sec:implementation}); and each coin is the debt of its issuer only: past holders bear no liability for it.

\mypara{Payment and state channels}
Layer-2 scaling solutions reduce but do not eliminate reliance on blockchain consensus~\cite{gudgeon2020layer2}.  Payment channels such as the Lightning Network~\cite{poon2016bitcoin} enable off-chain bilateral transactions but fall back to the blockchain for dispute resolution.  State channels~\cite{dziembowski2018general} generalise this to arbitrary state transitions.  Boyen et al.~\cite{boyen2016blockchain} proposed blockchain-free cryptocurrencies using a DAG structure, a prior attempt at the same design goal as grassroots currencies.  The abstract transactions of grassroots currencies are bilateral and resemble state channel updates in structure, but differ: state channels derive their security from the ability to publish the latest state on-chain in case of dispute, while grassroots currencies derive theirs from the participants' execution and the issuer's countersignature, with no blockchain fallback.  The voluntary swap of grassroots coins is related to cross-chain atomic swaps~\cite{herlihy2018atomic}, which exchange assets across two blockchains with no trusted third party; grassroots currencies achieve the same atomicity between the two parties, without any blockchain. Discreet Log Contracts~\cite{dryja2017discreet} are bilateral derivatives on Bitcoin whose payout is fixed by an oracle signature, with the terms of the contract invisible on chain; grassroots currencies share the bilateral form and settle with no chain---a transaction is a swap between the two parties, enforced by the written obligations of the coins rather than by a spending transaction.  The generalised state-channel frameworks~\cite{dziembowski2018general,coleman2018counterfactual} run arbitrary contracts off chain between two parties and keep the chain as the adjudicator of last resort; a grassroots coin has no such adjudicator, and a party who disowns their signature is in default, remedied at law and made public by the record (Section~\ref{sec:insolvency}).

\mypara{Monetary theory}
Grassroots currencies create \emph{inside money}~\cite{cavalcanti1999inside,cavalcanti1999model}---a medium of exchange backed by private credit rather than sovereign authority.  Cavalcanti and Wallace~\cite{cavalcanti1999model} studied a population of bankers and nonbankers: bankers' financial histories are public, enabling punishment for deviation, while nonbankers' histories are private, making notes an essential medium of exchange.  In grassroots currencies, everyone is between a banker and a nonbanker: financial records are not public, but the holder of another person's coins can choose which coin to redeem against it among those held by the other person, requiring the coin holder to have access to such knowledge.

Kocherlakota's result that ``money is memory''~\cite{kocherlakota1998money} is directly relevant: grassroots currencies replace monetary tokens with cryptographic records of bilateral obligations, with agents holding the distributed memory that makes money possible.  The Bank of England's account of how commercial banks create money through lending~\cite{mcleay2014money} describes a process structurally analogous to the opening of a mutual credit line: in both cases, the act of lending simultaneously creates both an asset (the loan) and a liability (the deposit or coin).  The theoretical foundation for such endogenous money creation through credit is established by Palley~\cite{palley2002endogenous}.  Brunnermeier et al.~\cite{brunnermeier2019digitalization} provide a comprehensive framework connecting inside/outside money, token/account forms, and digital currency competition, within which grassroots currencies occupy a distinctive position as peer-issued, record-based inside money operating without either sovereign backing or blockchain consensus.  Finally, Mundell's theory of optimum currency areas~\cite{mundell1961theory} is relevant to the regional bank and currency recognition mechanisms of Appendix~\ref{sec:community-bank}, which allow grassroots currency areas to form and merge bottom-up rather than top-down.

\mypara{Motivation and limitations}
Grassroots currencies are for communities in which people already trust each other---families, neighbours, colleagues, a community bank and its members---as a credit line is opened only between persons willing to hold each other's debt.  Among strangers no such line is opened, so the currencies form where trust already is, and reach further along the chains of Section~\ref{sec:coins}.  Without a shared ledger no person sees total issuance: each sees the coins they hold, and the holdings that friends disclose to each other, from which the ratio of Section~\ref{sec:ratio} is computed.  An estimate of a counterparty's default is thus as good as their disclosure.  A currency per person makes the number of currencies the number of persons; the holdings of each person are the currencies of the few they trade with, so the record a person keeps grows with their own circle and not with the economy.

\end{document}